\documentclass{new_tlp}
\usepackage{mathptmx}

\usepackage{amsmath}
\usepackage{graphicx}
\usepackage{multirow}
\usepackage{hyperref}
\usepackage{amssymb}
\usepackage{tikz}
\usepackage{pgfplots}
\usepackage{float}
\usepackage{svg}
\usepackage{multirow}
\usetikzlibrary{positioning}
\usepackage{fancyvrb}
\usepackage{listings}
\usepackage{lineno}

\newcommand{\myimplies}{\rightarrow}
\newcommand{\Next}{\textsf{X}}
\newcommand{\Eventually}{\textsf{F}}
\newcommand{\Always}{\textsf{G}}
\newcommand{\WNext}{\textsf{X$\sf _{\tiny w}$}}
\newcommand{\Until}{\textsf{U}}
\newcommand{\WUntil}{\textsf{W}}
\newcommand{\Release}{\textsf{R}}
\newcommand\Template[1]{$#1(a,b)$}
\usepackage{xcolor}
\usepackage{xspace}

\newcommand{\LTLf}{LTL$_\text{f}$\xspace}
\newcommand{\LTLp}{LTL$_\text{p}$\xspace}
\newcommand{\declare}{\textsf{Declare}\xspace}
\newcommand{\dpy}{\textsf{D4Py}\xspace}
\def\repository{https://github.com/ainnoot/padl-2024}
\def\repositoryurl{\href{https://github.com/ainnoot/padl-2024}{https://github.com/ainnoot/padl-2024}}
\usepackage{booktabs}
\newcommand{\myparagraph}[1]{\smallskip\noindent\textbf{#1.}\xspace}

\newtheorem{example}{Example}[section]  %

\definecolor{bluekeywords}{rgb}{0.13, 0.13, 1}
\definecolor{greencomments}{rgb}{0, 0.5, 0}
\definecolor{redstrings}{rgb}{0.9, 0, 0}
\definecolor{graynumbers}{rgb}{0.5, 0.5, 0.5}
\definecolor{asp-keyword}{RGB}{0,0,128}
\definecolor{asp-comment}{RGB}{0,128,0}
\definecolor{srr}{RGB}{255,0,0}
\lstdefinelanguage{ASP}{
  sensitive=true,
  breaklines=true,
  captionpos=b,
  showspaces=false,
  showstringspaces=false
}

\usepackage{listings}
\newcommand\bcmdtab{\noindent\bgroup\tabcolsep=0pt%
  \begin{tabular}{@{}p{10pc}@{}p{20pc}@{}}}
\newcommand\ecmdtab{\end{tabular}\egroup}

\title[Theory and Practice of Logic Programming]
    {Direct Encoding of Declare Constraints in ASP%
    \footnote[3]{
    Extended version of a distinguished paper of PADL'24~\cite{DBLP:conf/padl/ChiarielloFIR24} invited for rapid publication on TPLP.}}

\author[F. Chiariello, V. Fionda, A. Ielo and F. Ricca]{
FRANCESCO~CHIARIELLO$^{1}$,VALERIA~FIONDA$^{2}$, ANTONIO~IELO$^{2}$, FRANCESCO~RICCA$^{2}$\\
$^1$IRIT, ANITI, University of Toulouse, France\\
\email{francesco.chiariello@irit.fr}\\
$^2$Università della Calabria, Rende, Italia\\
\email{\{valeria.fionda,antonio.ielo,francesco.ricca\}@unical.it}
}

\jdate{March 2003}
\pubyear{2003}
\pagerange{\pageref{firstpage}--\pageref{lastpage}}
\doi{S1471068401001193}

\begin{document}

\label{firstpage}

\maketitle
\begin{abstract}
Answer Set Programming (ASP), a well-known declarative logic programming paradigm, has recently found practical application in Process Mining. In particular, ASP has been used to model tasks involving declarative specifications of business processes. In this area, \declare stands out as the most widely adopted declarative process modeling language, offering a means to model processes through sets of constraints valid traces must satisfy, that can be expressed in Linear Temporal Logic over Finite Traces (\LTLf). Existing ASP-based solutions encode \declare constraints by modeling the corresponding \LTLf formula or its equivalent automaton which can be obtained using established techniques. In this paper, we introduce a novel encoding for \declare constraints that \textit{directly} models their semantics as ASP rules, eliminating the need for intermediate representations. We assess the effectiveness of this novel approach on two Process Mining tasks by comparing it with alternative ASP encodings and a Python library for \declare. Under consideration in Theory and Practice of Logic Programming (TPLP).
\end{abstract}

\begin{keywords}
Answer Set Programming, Process Mining
\end{keywords}

\section{Introduction}
In the context of process mining, a \textit{process} typically refers to a sequence of events or activities that are performed in order to accomplish a particular goal within a business or organizational setting~\cite{pmdef}. Process Mining~\cite{DBLP:books/sp/22/Aalst22} is an interdisciplinary field offering techniques and tools to discover, monitor, and improve real processes by extracting knowledge from event logs readily available in today's information systems. One of the main tasks of Process Mining is \emph{conformance checking}, assessing the correctness of a specific execution of a process, known as \emph{trace}, against a \emph{process model}. Such a process model is a formal mathematical representation enabling various analytical and reasoning tasks related to the underlying process. Process models adopt either an imperative or a declarative language form, with the former explicitly describing all possible process executions, and the latter using logic-based constraints to define what is not permitted within process executions. Imperative process models are suitable for well-structured routine processes but often fall short in scenarios involving complex coordination patterns. Declarative models, conversely, offer a flexible alternative in such scenarios.

Linear temporal logic over finite traces (\LTLf)~\cite{DBLP:conf/ijcai/GiacomoV13} has emerged has a natural choice for declarative process modeling. However, within real-world Process Mining scenarios, \LTLf formulae that are unrestricted in structure are rarely used. Rather, a specific collection of predefined patterns, which have their origins in the domain of system verification~\cite{DBLP:conf/icse/DwyerAC99}, is commonly used. Limiting declarative modeling languages to a set of predefined patterns has two advantages: first, it simplifies modeling tasks~\cite{DBLP:conf/fm/GreenmanPSZGKMNZ24,DBLP:journals/programming/GreenmanSNK23}, and second, it allows for the development of specialized solutions that may outperform generic \LTLf reasoners in terms of efficiency. Specifically, \declare~\cite{DBLP:journals/ife/AalstPS09} is the declarative process modeling language most commonly used in Process Mining applications and consists of a set of patterns, referred to as {\textit{templates}}. The semantics of \declare templates is provided in terms of \LTLf, thus enabling logical reasoning and deduction processes within the \declare framework~\cite{DBLP:books/sp/22/CiccioM22}.

Recent research proposals suggest that Answer Set programming (ASP)~\cite{DBLP:journals/ngc/GelfondL91} can be successfully used for various tasks within declarative process mining~\cite{DBLP:conf/ijcai/IeloRP22,DBLP:conf/aaai/ChiarielloMP22}. Nevertheless, to the best of our knowledge, there has been no prior attempt to encode the \declare \LTLf patterns library using ASP in a \textit{direct} way. Here, ``direct'' means an encoding that captures the semantics of \declare constraints without the need for any intermediary translation. This paper fills this gap by introducing a direct encoding approach for the main \declare patterns. The proposed technique is benchmarked against existing ASP-based encodings across various logs commonly used in Process Mining~\cite{DBLP:conf/bpm/LopesF19}. The goal of the experimental analysis is twofold: first, to assess the performance of ASP-based methods in tasks related to conformance and query checking; and second, to evaluate the effectiveness of our proposed direct encoding strategy. To facilitate reproducibility, code and data used in our experiments are openly accessible at \texttt{\href{\repository}{\repository}}.

\myparagraph{Related Work} %
Answer Set Programming (ASP)~\cite{DBLP:journals/cacm/BrewkaET11,DBLP:journals/ngc/GelfondL91,DBLP:journals/amai/Niemela99} has shown promise in planning  to inject domain-dependent knowledge rules, resembling predefined patterns, encoded via an action theory language into ASP~\cite{mcilrath}. Researchers additionally explored injecting temporal knowledge, expressed as LTL formulae, in answer set planning~\cite{mcilrath}. An extension of ASP with temporal operators was proposed by \cite{DBLP:conf/lpnmr/CabalarKMS19}; and other applications of logic programming to solve %
process mining tasks have been developed in earlier works~\cite{DBLP:journals/topnoc/ChesaniLMMRS09,DBLP:conf/ilp/LammaMRS07}.
The works in which ASP has been used to tackle various computational tasks in \declare-based Process Mining~\cite{DBLP:conf/ijcai/IeloRP22} are closer to this paper. In ~\cite{DBLP:conf/aaai/ChiarielloMP22} authors propose a solution based on the well-known \LTLf-to-automata translation~\cite{DBLP:conf/ijcai/GiacomoV13}. 
In ~\cite{DBLP:conf/bpm/KuhlmannCG23}, the authors suggested a method for encoding the semantics of temporal operators into a logic program, enabling the encoding of arbitrary \LTLf formulae, by a reification of their syntax tree. Some other  recent studies~\cite{ChesaniFGGLMMMT22,ChesaniFGLMMMT23} tackle a slightly different objective by using ASP to directly encode \declare constraints but with the focus of distinguishing between normal and deviant process traces.

\myparagraph{Paper structure}{
The paper is organized as follows. Section 2 provides preliminaries on Process Mining, Linear Temporal Logic over Finite Traces (\LTLf) and Answer Set Programming (ASP); Section 3 adapts automata-based~\cite{DBLP:conf/aaai/ChiarielloMP22} and syntax tree-based~\cite{DBLP:conf/bpm/KuhlmannCG23} ASP encodings to \declare; Section 4 introduces our novel direct ASP encoding for \declare; Section 5 reports the results of the experimental evaluation. Section 6 concludes the paper.
}

\section{Preliminaries}
In this section fundamental concepts related to Process Mining, linear temporal logic over finite traces, the \declare process modeling language, and Answer Set Programming are discussed.

\subsection{Process Mining}
\emph{Process Mining}~\cite{DBLP:books/sp/22/Aalst22} is a research area at the intersection of Process Science and Data Science. It leverages data-driven techniques to extract valuable insights from operational processes by analyzing event data (i.e., event logs) collected during their execution. A process can be seen as a sequence of activities that collectively allow to achieve a specific goal. A trace represents a concrete execution of a process recording the exact sequence of events and decisions taken in a specific instance. Process Mining plays a significant role in Business Process Management~\cite{DBLP:books/sp/Weske19}, by providing data-driven approaches for the analysis of events logs directly extracted from enterprise information systems. 
Typical Process Mining tasks include: \emph{Conformance checking} that aims at verifying if a trace is conformant to a specified model and, for logic-based techniques, \emph{Query Checking} that evaluates \emph{queries} (i.e., formulae incorporating variables) against the event log.
Several formalisms can be used in process modelling, with Petri nets~\cite{DBLP:journals/jcsc/Aalst98} and BPMN~\cite{DBLP:journals/csi/ChinosiT12} being among the most widely used, both following an imperative paradigm. Imperative process models explicitly describe all the valid process executions and can be impractical when the process under consideration is excessively intricate. In such cases, declarative process modelling~\cite{DBLP:books/sp/22/CiccioM22} is a more appropriate choice. Declarative process models specify the desired properties (in terms of constraints) that each valid process execution must satisfy, rather than prescribing a step-by-step procedural flow. Using declarative modeling approaches allows to easily specify the desired behaviors: everything that does not violate the rules is allowed.
Declarative specifications are expressed in \declare~\cite{DBLP:journals/ife/AalstPS09}, LTL over Finite Traces (\LTLf)~\cite{DBLP:journals/fmsd/FinkbeinerS04}, or LTL over Process Traces (\LTLp)~\cite{DBLP:journals/jair/FiondaG18}. 

\subsection{Linear Temporal Logic over Finite Traces}
This section recaps minimal notions of Linear Temporal Logic over Finite Traces (\LTLf) \cite{DBLP:conf/ijcai/GiacomoV13}. We introduce finite traces, the logic's syntax and semantics, then informally describe its temporal operators, and some Process Mining application-specific notation.

\myparagraph{Finite Traces} Let $\mathcal{A}$ be a set of propositional symbols. A \emph{finite trace} is a sequence $\pi = \pi_0 \cdots \pi_{n-1}$ of $n$ propositional interpretations $\pi_i \subseteq \mathcal{A}$, $i = 0,\dots,n-1$, for some $n \in \mathbb{N}$. The interpretation $\pi_i$ is also called a \emph{state}, and $|\pi| = n$ denotes the trace \emph{length}.

\myparagraph{Syntax}  \LTLf is an extension of propositional logic that can be used to reason about temporal properties of traces. It shares the same syntax as Linear Temporal Logic (LTL) \cite{DBLP:conf/focs/Pnueli77}, but it is interpreted over \emph{finite} traces rather than \emph{infinite} ones. An \LTLf formula $\varphi$ over $\mathcal{A}$ is defined according to the following grammar:
\[
\varphi::= \top \mid a \mid \neg \varphi \mid \varphi \land \varphi %
\mid \Next \varphi %
\mid \varphi_1  \Until \varphi_2,
\]

\noindent where $a \in \mathcal{A}$. We assume common propositional ($\lor$, $\myimplies$, $\longleftrightarrow$, etc.) and temporal logic shorthands. In particular, for temporal operators, we define the \emph{eventually} operator $\Eventually \varphi \equiv \top \Until \varphi$, the \emph{always} operator $\Always \varphi \equiv \lnot \Eventually \lnot \phi$, the \emph{weak until} operator $\varphi \WUntil \varphi' \equiv \Always\varphi \lor \varphi \Until \varphi'$ and \emph{weak next} operator $\WNext \varphi \equiv \lnot \Next \lnot \varphi \equiv \Next \varphi \ \lor \lnot \Next \top$.

\myparagraph{Semantics} Let $\varphi$ be an \LTLf formula, $\pi$ a finite trace, $0 \le i < |\pi|$ an integer. The \emph{satisfaction} relation, denoted by $\pi, i \models \varphi$, is defined recursively as follows:

\begin{itemize}
\item[$\bullet$] $\pi, i \models \top$; %
\item[$\bullet$] $\pi,i\models a$ iff  $a \in \pi_i$;
\item[$\bullet$] $\pi, i\models \neg \varphi$ iff $\pi,i \models \varphi$ does not hold;%
\item[$\bullet$] $\pi, i\models \varphi_1 \wedge \varphi_2$ iff $\pi,i\models \varphi_1$ and $\pi,i\models \varphi_2$;
\item[$\bullet$] $\pi,i\models \Next \varphi$  iff   $i<|\pi|-1$ and $\pi,{i+1}\models\varphi$;
\item[$\bullet$] $\pi,i\models \varphi_1 \Until \varphi_2$ iff $\exists j$ with $i\leq j\leq |\pi|-1$ s.t. $\pi,j\models \varphi_2$ and $\forall k$ with $i\leq k<j$, $\pi,k\models \varphi_1$.
\end{itemize}

We say that $\pi$ is a \emph{model} for $\varphi$ if $\pi, 0 \models \varphi$, denoted as $\pi \models \varphi$. %
{Although \LTLf and LTL share the same syntax, interpreting a formula over finite traces results in very different properties. As an example, consider the \emph{fairness} constraint of the form $\Always \Eventually \varphi$. In LTL, this kind of formulae means that \textit{it is always true that in the future $\varphi$ holds}. However, in \LTLf, this is equivalent to stating that $\varphi$ \textit{holds in the last state of the trace}. Thus, while the formula admits only infinite traces as counterexamples when interpreted in LTL, it admits finite counterexamples when interpreted as \LTLf. It holds more generally that, as stated in ~\cite{DBLP:conf/ijcai/GiacomoV13},  \textit{direct nesting of temporal operators yields un-interesting formulae in \LTLf}.}

\myparagraph{Automaton associated to \LTLf formulae} Each \LTLf formula $\varphi$ over $\mathcal{A}$ can be associated to a minimal finite-state automaton $\mathcal{M}(\varphi)$ over the alphabet $2^\mathcal{A}$ such that for any trace $\pi$ it holds that $\pi \models \varphi$ iff $\pi$, is accepted by $\mathcal{M}(\varphi)$~\cite{DBLP:conf/aips/GiacomoF21,DBLP:conf/ijcai/GiacomoV13}.
A common assumption in \LTLf applications to Process Mining, referred to as \emph{\declare assumption}~\cite{DBLP:conf/aaai/GiacomoMM14} or \emph{simplicity assumption}~\cite{DBLP:conf/bpm/ChiarielloPM23}, is that exactly one activity occurs in each state.  \LTLf with this additional restriction is known as \LTLp~\cite{DBLP:journals/jair/FiondaG18}.
The assumption has the following practical implication: given a \LTLp formula $\varphi$, the minimal automaton $\mathcal{M}(\varphi)$ of $\varphi$ can be simplified into a deterministic automaton over $\mathcal{A}$~\cite{DBLP:conf/bpm/ChiarielloPM23}, as shown in Figure~\ref{fig:response_cugini}.

\begin{figure}[t!]
\centering
\begin{minipage}{\linewidth}
\begin{minipage}{0.32\linewidth}
\centering
\begin{figure}[H]
\includesvg[width=\linewidth]{figures/response_ltlf.svg}
\end{figure}
\end{minipage}
\hspace{0.25\linewidth}
\begin{minipage}{0.32\linewidth}
\centering
\begin{figure}[H]
\includesvg[width=\linewidth]{figures/response_ltlp.svg}
\end{figure}
\end{minipage}
\end{minipage}
\caption{Left: Minimal automaton for the \LTLf formula $\varphi = \Always(a \rightarrow \Next\Eventually b)$. Models (labeling the transitions) represent sets of symbols; Right: Minimal automaton for $\varphi$ interpreted as a \LTLp formula, where $*$ denotes any $x \in \mathcal{A} \setminus \{a, b\}$. A comma on edges denotes multiple transitions. }
\label{fig:response_cugini}
\end{figure}

\begin{table}[t]
    \centering    
    \resizebox{\linewidth}{!}{
    \begin{tabular}{rc}
        \hline
        \textbf{Template} & \textbf{\LTLp}\\
        \hline
        \Template{Choice} & $\Eventually(a \lor b)$\\
        \Template{ExclusiveChoice} & $\Template{Choice} \land \neg (\Eventually a \land \Eventually b)$\\
        \Template{RespondedExistence} & $\Eventually a \myimplies \Eventually b $\\
        \Template{Coexistence} & $\Template{RespondedExistence}$$ \land \textsf{RespondexExistence(b,a)}$\\
        \Template{Response} & $\Always (a \myimplies \Eventually b)$\\
        \Template{Precedence} & $\neg b \ \WUntil \ a$\\
        \Template{Alt.Response} & $\Always(a \myimplies $$ \Next(\neg a \Until b))$\\
        \Template{Alt.Precedence} & $\Template{Precedence} \land \Always(b \myimplies \textcolor{red}{\WNext} \Template{Precedence})$\\
        \Template{ChainResponse} & $\Always(a \myimplies \Next b)$\\
        \Template{ChainPrecedence} & $\Always(\Next b \myimplies a) \textcolor{red}{\land\neg b}$\\
        \hline
    \end{tabular}}\vspace*{0.01cm}
    \hspace{1em}
        \caption{Some \declare templates as \LTLp formulae. We slightly edit the definitions for \Template{ChainPrecedence} and \Template{AlternatePrecedence}, to align their semantics to the informal description commonly assumed in Process Mining applications. Changes w.r.t the original source are highlighted in red. The \textsf{Succession} (resp. \textsf{AlternateSuccession}, \textsf{ChainSuccession}) template is defined as the conjunction of (\textsf{Alternate}, \textsf{Chain}) \textsf{Response} and \textsf{Precedence} templates.
        }
    \label{tab:declare_templates}
\end{table}

\subsection{\declare modeling language} 
\declare~\cite{DBLP:journals/ife/AalstPS09} is a declarative process modeling language that consists of a set of \emph{templates} that express temporal properties of process execution traces. The semantics of each \declare template is defined in terms of an underlying \LTLp formula. Table~\ref{tab:declare_templates} provides the \LTLp definition of some \declare templates, as reported in \cite{DBLP:conf/aaai/GiacomoMM14}. \declare\ templates can be classified into four distinct categories, each addressing different aspects of process behavior: \textit{existence} templates, specifying the necessity or prohibition of executing a particular activity, potentially with constraints on the number of occurrences; \textit{choice} templates, centered around the concept of execution choices as they model scenarios where there is an option regarding which activities may be executed; \textit{relation} templates, establishing a dependency between activities as they dictate that the execution of one activity necessitates the execution of another, often under specific conditions or requirements; \textit{negation} templates, modelling mutual exclusivity or prohibitive conditions in activity execution. In Table~\ref{tab:declare_templates}, \Template{Choice} and \Template{ExclusiveChoice} are examples of \textit{choice} templates; while the others fall under the \textit{relation} category.
{A \declare model is a set of \emph{constraints}, where a constraint is a particular instantiation of a template, over specific activities, called respectively \emph{activation} and \emph{target} for binary constraints.} Informally, the activation of a \declare constraint is the activity whose occurrence imposes a constraint over the occurrence of the target on the rest of the trace. A more formal account of activation-target semantics of \declare constraints can be found in~\cite{DBLP:conf/bpm/CecconiCGM18}. 

\begin{figure}[t]
\includesvg[width=.8\linewidth]{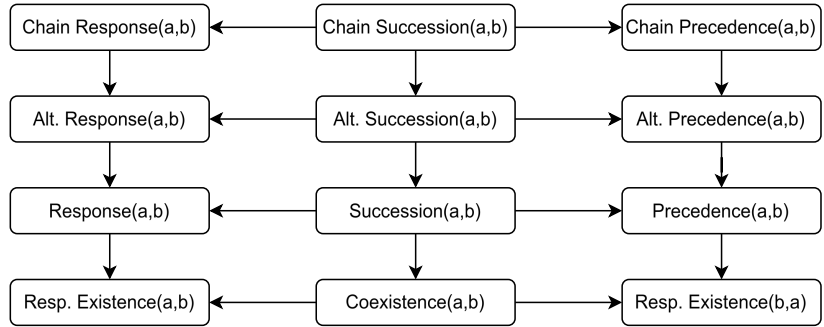}
\caption{\textit{Subsumption hierarchy} between \declare relation and existence templates. The arrow $t_1 \rightarrow t_2$ denotes that $\pi \models t_1 \implies \pi \models t_2$, e.g. $t_1$ is more specific (\textit{subsumes}) than $t_2$.}
\label{fig:subsumption_hierarchy}
\end{figure}

\myparagraph{Main \declare constraints} {\declare constraints are arranged into ``chains'', that strengthen/weaken the basic relation templates $Response(a,b)$ and $Precedence(a,b)$. The $Response(a,b)$ constraint specifies that whenever $a$ occurs, $b$ must occur after, while the $Precedence(a,b)$ constraint states that $b$ can occur only if $a$ was executed before. The constraints $AlternateResponse(a,b)$ and $AlternatePrecedence(a,b)$ strengthen respectively the $Response(a,b)$ and $Precedence(a,b)$ constraints, imposing that involved activities must ``alternate'', e.g. once the activation occurs, the target must occur before the activation re-occurs. The constraint $ChainResponse(a,b)$ imposes that the target must immediately follow the activation (that is, $\pi_t = a \rightarrow \pi_{t+1} = b$); the constraint $ChainPrecedence(a,b)$ that the target must immediately precede the activation (that is, $\pi_t = b \rightarrow \pi_{t-1} = a$).}
{We can also weaken the temporal relation properties: the $Coexistence$ relation states that two activities must occur together; the $ExclusiveChoice$ states that exactly one of them must occur; $RespondedExistence$ that one implies the occurrence of the other. These templates do not specify temporal behavior, but only co-occurrences of events. Finally, the $Succession$ template family combines, by means of propositional conjunction, $Response$ and $Precedence$ constraints.}
This hierarchy of constraints, graphically represented in Figure~\ref{fig:subsumption_hierarchy}, suggests a progression from more general properties, at the bottom of the hierarchy, to more specific ones, at the top. As an example, the $ChainResponse(a,b)$ is the most specific constraint in the $Response$ chain; it implies $AlternateResponse(a,b)$, which in turn implies $Response(a,b)$, which implies $RespondedExistence(a,b)$.

The following example showcases the informal semantics of the Response template, which will serve as a running example in the rest of the paper in the ASP encodings section.

{
\begin{example}[Semantics of the Response template]
The informal semantics for \Template{Response} is that \emph{whenever $a$ occurs in the trace, $b$ will appear in the future}. Formally, the template is defined as $\Always(a \myimplies \Eventually b)$. Thus, if $a$ occurs at time $t$ in a trace $\pi$, for the constraint to be satisfied, $b$ must appear in the trace suffix $\pi_{t+1}, \dots, \pi_n$. In the context of a customer service process, let's consider the Response template instantiated with $\textsf{a}=\textsf{customer\_complains}$ and $\textsf{b}=\textsf{address\_complain}$, corresponding to the template instantiation, i.e., the constraint, \textsf{Response(customer\_complains,} \textsf{address\_complain)}. Such constraint imposes that when a customer complaint is received (activation activity), a follow-up action to address the complaint (target activity), must be executed. On the other hand, the trace $\pi = (\textsf{customer\_complains},$ $\textsf{logging\_complain},$ $\textsf{address\_complain},$ $\textsf{feedback}\_$ $\textsf{collection})$ satisfies the above constraint. In contrast, the trace $\pi' = \textsf{customer\_complains},$ $\textsf{logging\_complain},$ $\textsf{address\_complain},$ $\textsf{customer\_complains},$ $\textsf{feedback\_collection}$ does not since the second occurrence of \textsf{customer\_complains} is not followed by any \textsf{address\_complain} event.
\end{example}
}

Another example shows how the $Response$ constraint can be further specialized.

{
\begin{example}[Semantics of the $AlternateResponse$ and $ChainResponse$ templates]\label{ex:cf}
Consider the execution traces $\pi_1 = aaabc$, $\pi_2 = abacb$, $\pi_3 = abab$. The constraint $Response(a,b)$ is valid over all these traces, while $\pi_1$ violates $AlternateResponse(a,b)$ because $a$ repeats before $b$ occurs after the first occurrence of $a$; The constraint $ChainResponse(a,b)$ is valid only on trace $\pi_3$, since both $\pi_1$ and $\pi_2$ have an $a$ that is not immediately followed by a $b$.
\end{example}}

\myparagraph{Conformance and Query Checking Tasks} 
This paper focuses on the \declare \emph{conformance checking} and \declare \emph{(template) query checking} tasks, as defined below:

Let $\mathcal{L}$ be an event log (a multiset of traces) and $\mathcal{M}$ a \declare model. The \textit{conformance checking} task $(\mathcal{L}, \mathcal{M})$ consists in computing the subset of traces $\mathcal{L}' \subseteq \mathcal{L}$ such that for each $\pi \in \mathcal{L}'$, $\pi \models c$ for all $c \in \mathcal{M}$. Basically, conformance checking determines whether a give process execution is compliant with a process model. 

Let $\mathcal{L}$ be an event log, and $c$ a constraint. {The support of $c$ on $\mathcal{L}$, denoted by $\sigma(c,\mathcal{L})$, is defined as the fraction of traces $\pi \in \mathcal{L}$ such that $\pi\models c$.} High support for a constraint is usually interpreted as a measure of relevance for the given constraint on the log $\mathcal{L}$. Given a \declare template $t$ and a \emph{support threshold} $s \in (0,1]$, the \textit{query checking} task $(t, \mathcal{L}, s)$ consists in computing variable-activity bindings such that the constraint $c$ we obtain by instantiating $t$ with such bindings has a support greater than $s$ on $\mathcal{L}$. {Query checking enables to discover which instantiation of a given template exhibit high (above the threshold) support on an event log, and is a valuable tool to explore and analyze event logs~\cite{DBLP:conf/otm/RaimCMMM14}.}

{
\begin{example} [Query checking and conformance checking]
Consider the template $Response(a, ?y)$ {(where $?y$ is a placeholder for an activity)} over the log $L = \{abab, abac, abadabd\}$, and with support $\sigma=0.5$. All the possible instantiations of the ``partial template'', which we obtain by substituting the placeholder $?y$ with an activity that appears in the event log, are the constraints $Response(a,b)$, $Response(a,c)$, and $Response(a,d)$, which yield respectively support $1, \frac{1}{3}, \frac{2}{3}$ over $L$. The substitution $?y = a$, yields an unsatisfiable \LTLf formula, hence zero support. Thus, the query checking task $(Response(a, ?y), L, \sigma=0.5)$ admits $\{Response(a,b), Response(a,d)\}$ as answers. In fact, if we perform the conformance checking task of the constraint $Response(a,c)$ the only compliant trace would be $abac$, with a support below 0.5.
\end{example}}

Interested readers can refer to ~\cite{DBLP:books/sp/22/CiccioM22} as a starting point for \declare-related literature. 

\subsection{Answer Set Programming}  
Answer set programming (ASP)~\cite{DBLP:journals/cacm/BrewkaET11,DBLP:journals/ngc/GelfondL91}
is a declarative programming paradigm based on the stable models semantics, which has been used to solve many complex AI problems~\cite{DBLP:journals/aim/ErdemGL16}.
We now provide a brief introduction describing the basic language of ASP. We refer the interested reader to~\cite{DBLP:journals/cacm/BrewkaET11,DBLP:journals/ngc/GelfondL91,DBLP:series/synthesis/2012Gebser} for a more comprehensive description of ASP. 
The syntax of ASP follows Prolog's conventions: variable terms are strings starting with an uppercase letters; constant terms are either strings starting by lowercase letter or are enclosed in quotation marks, or are integers. An \emph{atom} of arity $n$ is an expression of the form $p(t_1,\ldots,t_n)$ where $p$ is a predicate and $t_1,\ldots,t_n$ are terms. 
A (positive) \emph{literal} is an atom $a$ or its negation (negative literal) $not\ a$ where $not$ denotes negation as failure.
A \emph{rule} is an expression of the form $h :\!-\ b_1,\ldots,b_n$ where $b_1, \ldots, b_n$ is a conjunction of literals, called the \emph{body}, $n\geq 0$, and $h$ is an atom
called the \emph{head}. All variables in a rule must occur in some positive literal of the body. 
A \emph{fact} is a rule with an empty body (i.e., $n=0$). 
A \emph{program} is a finite set of rules.
Atoms, rules and programs that do not contain variables are said to be ground.
The Herbrand Universe $U_P$ is the collection of constants in the program $P$. 
The Herbrand Base $B_P$ is the set of ground atoms that can be generated by combining predicates from $P$ with the constants in $U_P$.
The ground instantiation of $P$, denoted by $ground(P)$, is the union of ground instantiations of rules in $P$ that are obtained by replacing variables with constants in $U_P$. 
An \emph{interpretation} $I$ is a subset of $B_P$. 
A positive (resp. negative) literal $\ell$ is true w.r.t. $I$, if $\ell \in I$ (resp. $\overline{\ell} \notin I$); it is false  w.r.t. $I$ if $\ell \notin I$ (resp. $\overline{\ell} \in I$). 
An interpretation $I$ is a \emph{model} of $P$ if for each $r \in ground(P)$, the head of $r$ is true whenever the body of $r$ is true.
Given a program $P$ and an interpretation $I$, the (Gelfond-Lifschitz) reduct~\cite{DBLP:journals/ngc/GelfondL91} $P^I$ is the program obtained from $ground(P)$ by (i)~removing all those 
rules having in the body a false negative literal w.r.t. $I$, and (ii)~removing negative literals from the body of remaining rules. 
Given a program $P$, the model $I$ of $P$ is a \emph{stable model} or \emph{answer set} if there is no $I^{\prime} \subset I$ such that $I^{\prime}$ is a model of $P^I$. 

In the paper, also use more advanced ASP constructs such as \emph{choice rules} and \textit{function symbols}. {A choice rule is an expression of the form 
$$L \ \{h_1; \dots; h_k\} \ U :\!-\ b_1, \dots, b_n. $$
where $h_i$ are atoms and $b_i$ are literals. Its semantics is that whenever the body of the rule is satisfied in an interpretation $I$, then $L \le |I \cap \{h_1, \dots, h_k\}| \le U$. Choice rules can be rewritten into a set of normal rules~\cite{DBLP:series/synthesis/2012Gebser}.}

{
A function symbol is a ``composite term'' of the form $f(t_1, \dots, t_k)$ where $t_i$ are terms and $f$ is a predicate name. For what we are concerned in this paper, function symbols will be essentially used to model the input, acting as syntactic sugar that simplifies the modeling of records. This presentation choice makes the encodings more readable, since they allow for a compact notation of records. It is easy to see that function symbols can be easily replaced by more lengthy standard ASP specifications with additional terms in the input predicates.
}

We refer the reader to~\cite{DBLP:journals/tplp/CalimeriFGIKKLM20} for a description of more advanced ASP constructs. In the rest of the paper, ASP code examples will use \textsc{Clingo}~\cite{clingo} syntax. 

\section{Translation-based ASP encodings for \declare}
This section introduces ASP encodings for conformance checking of \declare models and query checking of \declare constraints with respect to an input event log, based on the translation to automata and syntax trees. Both encodings share the same input fact schema to specify which \declare constraints belong to the model, or which constraint we are performing query checking against. These encodings are \emph{indirect}, since they rely on a translation, but also \emph{general} in the sense that they can be applied to the evaluation of arbitrary \LTLp formulae. This is achieved, in the case of the syntax tree encoding, by reifying the syntax tree of a formula and by explicitly modeling the semantics of each \LTLp temporal operator through a logic program, and in the case of the automaton encoding, by exploiting the well-known \LTLp-to-automaton translation~\cite{DBLP:conf/bpm/ChiarielloPM23,DBLP:conf/ijcai/GiacomoV13}. Thus, one can use these two encodings to represent \declare constraints by their \LTLp definitions. The automaton-based encoding is adapted from~\cite{DBLP:conf/aaai/ChiarielloMP22}, the syntax tree-based encoding is adapted from~\cite{DBLP:conf/bpm/KuhlmannCG23,DBLP:conf/sum/KuhlmannC24} --- integrating changes to allow for the above-mentioned shared fact schema and evaluation over multiple traces. A similar encoding has also been used in~\cite{DBLP:conf/ilp/IeloLF23} to learn \LTLf formulae from sets of example traces (using the ASP-based inductive logic programming system ILASP~\cite{ilasp}) and to implement an ASP-based \LTLf bounded satisfiability solver~\cite{DBLP:conf/lpnmr/FiondaIR24}. 

{Earlier work in answer set planning~\cite{mcilrath} also exploited LTL constraints using a similar syntax tree-reification approach}. We start by defining how event logs and \declare constraints are encoded into facts, then introduce conformance checking and query checking encodings with the two approaches.

\subsection{Encoding process traces and \declare models}

\myparagraph{Encoding process traces} For our purposes, an event log $\mathcal{L}$ is a multiset of process traces, thus a multiset of strings over an alphabet of propositional symbols $\mathcal{A}$ (representing activities). We assume that each trace $\pi \in \mathcal{L}$ is uniquely indexed by an integer, and we denote that the trace $\pi$ has index $i$ by $id(\pi) = i$. This is a common assumption in Process Mining, where $i$ is referred to as the \emph{trace identifier}. Traces are modeled through the predicate \texttt{trace/3}, where the atom $trace(i, t, a)$ encodes that $\pi_t=a, id(\pi) = i$ --- that is, the $t$-th activity in the $i$-th trace $\pi$ is $a$. Given a process trace $\pi$, we denote by $E(\pi)$ the set of facts that encodes it. Thus, an event log $\mathcal{L}$ is encoded as $E(\mathcal{L}) = \bigcup_{\pi \in \mathcal{L}} E(\pi)$.

\begin{example}[Encoding a process trace]
Consider an event log composed of the two process traces $\pi^0 = abc$ and $\pi^1 = xyz$, respectively with identifiers 0 and 1, over the propositional alphabet $\mathcal{A} = \{a, b, c, x, y, z\}$. This is encoded by the following set of facts:

\begin{lstlisting}
trace(0,0,a). trace(0,1,b). trace(0,2,c). trace(1,0,x). trace(1,1,y). trace(1,2,z).
\end{lstlisting}

{In our encoding, activities are represented as constants that appear at least once in a trace in the event log, e.g. $a$ such that $trace(\_,\_,a)$ is a fact in the input event log.}

\end{example}

Each \declare template, informally, can be understood as a ``\LTLp formula with variables''. Substituting these variables with activities yields a \declare constraint. How templates are instantiated into constraints, and how constraints are evaluated over traces, depends on the ASP encoding we use. However, all encodings share a common fact schema where constraints are expressed as templates with bound variable substitutions.

\myparagraph{Encoding \declare constraints} A \declare constraint is modeled by predicates \texttt{constraint/2} and \texttt{bind/3}. The former model which \declare template a given constraint is instantiated from and the latter which activity-variable bindings instantiate the constraint. An atom $constraint(cid,$ $template)$ encodes that the constraint uniquely identified by $cid$ is an instance of the template $template$. The atom $bind(cid, arg, value)$ encodes that the constraint uniquely identified by $cid$ is obtained by binding the argument $arg$ to the activity $value$. Given a \declare model $\mathcal{M} = \{c_1, \dots, c_n\}$, where the subscript $i$ uniquely indexes the constraint $c_i$, we denote by $E(\mathcal{M})$ the set of facts that encodes $\mathcal{M}$, that is $E(\mathcal{M}) = \bigcup_{c \in \mathcal{M}} E(c)$. Recall that in \declare $\pi \models \mathcal{M}$ if and only if $\pi \models c$ for all $c \in \mathcal{M}$, thus there is no notion of ``order" among the constraints within $\mathcal{M}$ and it does not matter how indexes are assigned to constraints as long as they are unique.

\begin{example}[Encoding a \declare model] Consider the model $\mathcal{M}$ composed of the two constraints $\text{Response}(a_1, a_2)$ and $\text{Precedence}(a_2, a_3)$. $\mathcal{M}$ is encoded by the following facts:

\begin{minipage}{\linewidth}
\centering
\begin{minipage}{0.45\linewidth}
\begin{lstlisting}
constraint(0,"Response").
bind(0,arg_0,a_1). bind(0,arg_1,a_2).
\end{lstlisting}
\end{minipage}
\hspace{0.05\linewidth}
\begin{minipage}{0.45\linewidth}
\begin{lstlisting}
constraint(1,"Precedence").
bind(1,arg_0,a_2). bind(1,arg_1,a_3).
\end{lstlisting}
\end{minipage}
\end{minipage}
\end{example}

\subsection{Encoding Conformance Checking}

{All the \declare conformance checking encodings we propose consist of a {stratified normal logic program~\cite{DBLP:books/sp/Lloyd87,DBLP:books/mk/minker88/AptBW88}} $P_{CF}$ such that given a log $\mathcal{L}$ and a \declare model $\mathcal{M}$ we have that for all $\pi_i \in \mathcal{L}$, $\pi_i \models c_j \in \mathcal{M}$ if and only if the unique model of $P_{CF} \cup E(\mathcal{M}) \cup E(\mathcal{L})$ contains the atom $sat(i, j)$. The binary template $\text{Response}$ will be our example to showcase the different encodings.} Complete encodings for all the templates in Table \ref{tab:declare_templates} are available \href{\repository}{online}. 

\myparagraph{Automaton encoding} The automaton encoding, reported in Figure~\ref{fig:chiariello_enc}, models \declare templates through their corresponding automaton obtained by translating the template's \LTLp definition~\cite{DBLP:conf/aips/GiacomoF21}. The automaton's complete transition function is reified into a set of facts that defines the template in ASP. The predicates \texttt{initial/2, accepting/2} model the initial and accepting states of the automaton, while \texttt{template/4} stores the transition function of the template-specific automaton. In particular, \texttt{arg\_0} refers to the template \emph{activation}, and \texttt{arg\_1} refers to the template \emph{target}. A constraint $c$ instantiated from a template binds its \texttt{arg\_0}, \texttt{arg\_1} to specific activities. The constant \texttt{"*"} is used as a placeholder for any activity in $\mathcal{A} \setminus \{x, y\}$ -- where $x$ and $y$ are the bindings of \texttt{arg\_0} and \texttt{arg\_1}. %
Activities not explicitly mentioned as within the atomic propositions in an \LTLp formula $\varphi$ have the same influence to $\pi \models \varphi$. Consequently, all unbound activities can be denoted by the symbol \texttt{"*"} in the automaton transition table.
As an example, consider the constraint $c = \textsf{Response(a,b)}$, shown in Figure~\ref{fig:chiariello_example}. Evaluating the trace \texttt{abwqw} is equivalent to evaluating the trace \texttt{abtts}, which would be equivalent to evaluating the trace \texttt{ab***}, since $a$ and $b$ are the only propositional formulae that appear in the definition of $\textsf{Response(a,b)}$.

\begin{figure}
\begin{minipage}{\linewidth}
\centering
\begin{minipage}{0.45\linewidth}
\begin{lstlisting}

cur_state(C,TID,S,0) :- 
  trace(TID,_,_), 
  initial(Template,S), 
  constraint(C,Template).

last(TID,T) :- 
  trace(TID,T,_), 
  not trace(TID,T+1,_).

sat(TID,C) :- 
  cur_state(C,TID,S,T+1),
  last(TID,T),
  template(Template,C),
  constraint(C, Template),
  accepting(Template,S).

\end{lstlisting}
\end{minipage}
\hspace{0.05\linewidth}
\begin{minipage}{0.45\linewidth}
\begin{lstlisting}
cur_state(C,TID,S2,T+1) :- 
  cur_state(C,TID,S1,T), 
  constraint(C,Template), 
  template(Template,S1,Arg,S2), 
  trace(TID,T,A), 
  bind(C,Arg,A).



cur_state(C,TID,S2,T+1) :- 
  cur_state(C,TID,S1,T), 
  constraint(C,Template), 
  template(Template,S1,"*",S2), 
  trace(TID,T,A), 
  not bind(C,_,A).


\end{lstlisting}
\end{minipage}
\end{minipage}
\caption{ASP program to execute a finite state machine corresponding to a constraint, encoded as \texttt{template/4} facts, on input strings encoded by \texttt{trace/3} facts.}
\label{fig:chiariello_enc}
\end{figure}

\myparagraph{Syntax tree-based encoding} The syntax tree encoding, shown in Figure~\ref{fig:corea_enc}, reifies the syntax tree of a \LTLp formula into a set of facts, where each node represents a sub-formula. The semantics of temporal operators and propositional operators is defined in terms of ASP rules. Analogously to the automaton encoding, templates are defined in terms of reified syntax trees, which are used to evaluate each constraint according to the template they are instantiated from. The following normal rules define the semantics of each temporal and propositional operator. We report the rules for operators $\{\Until, \Next, \lnot, \land\}$ which are the basic operators of \LTLp. The full encoding, that also includes definitions of derived operators, is available \href{\repository}{online}. In particular, the \texttt{true/4} predicate tracks which sub-formula of a constraints' definition is true at any given time. As an example, the atom $true(c, f, t, i)$ encodes that \emph{at time $t$ the constraint sub-formula $f$ of constraint $c$ is satisfied on the $i$-th trace}. The terms \texttt{conjunction/3}, \texttt{negate/2}, \texttt{next/2}, \texttt{until/3} model the topology of the syntax tree of the corresponding formula. The first term refers to a node identifier, while the other terms (one for unary operators, two for the binary operators are the node identifiers of its child nodes. The \texttt{atom/2} predicate models that a given node (first term) is an atom, bound to a particular argument (second term) by the \texttt{bind/3} predicate which is used in encoding of \declare constraints. Figure~\ref{fig:corea_ex} shows an example.

\begin{figure}
\begin{minipage}{\linewidth}
\centering
\begin{minipage}{0.45\linewidth}
\begin{lstlisting}
last(TID,T) :- 
  trace(TID,T,_),
  not trace(TID,T+1,_).

true(C,F,T,TID) :-
  constraint(C,Template),
  template(Template, 
    atom(F,Arg)
  ),
  bind(C,Arg,A),
  trace(TID,T,A).

true(C,F,T,TID) :- 
  constraint(C,Template), 
  template(Template,
    conjunction(F,G,H)
  ), 
  trace(TID,T,_), 
  true(C,G,T,TID), 
  true(C,H,T,TID).

true(C,F,T,TID) :-
  constraint(C,Template), 
  template(Template,negate(F,G)), 
  not true(C,G,T,TID), 
  trace(TID,T,_).
\end{lstlisting}
\end{minipage}
\hspace{0.05\linewidth}
\begin{minipage}{0.45\linewidth}
\begin{lstlisting}
sat(C,TID) :- 
  true(C,0,0,TID).

true(C,F,Ti,TID) :-
  constraint(C,Template), 
  template(Template,next(F,G)), 
  trace(TID,Ti,_), 
  Tj=Ti+1, 
  Ti<M,
  last(TID,M), 
  true(C,G,Tj,TID).
  
true(C,F,Ti,TID) :- 
  constraint(C,Template), 
  template(Template, until(F,G,H)), 
  trace(TID,Ti,_), 
  trace(TID,Tj,_),
  Tj>=Ti, Tj<=M,
  last(TID,M),
  {true(C,G,T,TID): trace(TID,T,_), T>=Ti, T<Tj} = Tj-Ti,
  true(C,H,Tj,TID).
\end{lstlisting}
\end{minipage}

\end{minipage}
\caption{ASP program to evaluate each sub-formula of the \LTLp definition of a given template, encoded as \texttt{template/2} facts, on input strings encoded by a syntax tree representation through the \texttt{conjunction/3}, \texttt{negate/2}, \texttt{until/3}, \texttt{next/2} and \texttt{atom/2} terms.}
\label{fig:corea_enc}
\end{figure}

\begin{figure}
\begin{minipage}{\linewidth}
\begin{minipage}{0.45\linewidth}
\centering
\begin{lstlisting}
template("Response",0,"*",0).
template("Response",0,arg_1,0).
template("Response",0,arg_0,1).
template("Response",1,arg_1,0).
template("Response",1,"*",1).
template("Response",1,arg_0,1).
accepting("Response",0).
initial("Response",0).
\end{lstlisting}
\end{minipage}
\hspace{0.05\linewidth}
\begin{minipage}{0.38\linewidth}
\centering
\begin{figure}[H]
\includesvg[width=.8\linewidth]{figures/response_fsm.svg}
\end{figure}
\end{minipage}
\end{minipage}
\caption{Left: Facts that encode the Response template; Right: A minimal finite state machine whose recognized language is equal to the set of models of Response, under \LTLp semantics.}
\label{fig:chiariello_example}
\end{figure}

\begin{figure}
\begin{minipage}{\linewidth}
\centering
\begin{minipage}{0.45\linewidth}
\begin{lstlisting}
template("Response",always(0,1)).
template("Response",implies(1,2,3)).
template("Response",atom(2,arg_0)).
template("Response",eventually(3,4)).
template("Response",atom(4,arg_1)).
\end{lstlisting}
\end{minipage}
\hspace{0.08\linewidth}
\begin{minipage}{0.45\linewidth}
\begin{figure}[H]
\centering
\resizebox{!}{3cm}{
\begin{tikzpicture}
    \node[]                                     (arg02)         []      {$\text{arg}_0^2$};
    \node[]                                     (f)             [right=of arg02]    {$\textsf{F}^3$};
    \path (arg02.east) -- node[yshift=0.3cm]    (imp) {$\rightarrow^1$} (f.west);
    \node[yshift=-0.8cm]                        (G)             [above= of imp]                    {$\textsf{G}^0$};
    \node[yshift=0.8cm]                         (arg14)         [below=of f]        {$\text{arg}_1^4$};
    \draw[-] (G) -- (imp) node[] {};
    \draw[-] (imp) -- (arg02.east) node[] {};
    \draw[-] (imp) -- (f.west) node[] {};
    \draw[-] (f) -- (arg14) node[] {};
\end{tikzpicture}
}
\end{figure}
\end{minipage}
\end{minipage}
\caption{Left: Facts that encode the Response template; Right: Syntax tree of the Response template \LTLp definition. 
\label{fig:corea_ex}}
\end{figure}

\subsection{Encoding Query Checking}
 The query checking problem takes as input a \declare template $\mathcal{T}$, an event log $\mathcal{L}$ and consists in deciding which constraints $c$ can be instantiated from $\mathcal{T}$ such that $\sigma(c,\mathcal{L}) \ge k$, where $\sigma(c, \mathcal{L})$ is the support and denotes the fraction of traces in $\mathcal{L}$ that are models of $c$. The problem has been formally introduced in \cite{chan-temporal-logical-queries} for temporal logic formulae, and in \cite{DBLP:conf/otm/RaimCMMM14} it has been framed into a Process Mining setting, in the context of \LTLf. An ASP-based solution to the problem has been provided in \cite{DBLP:conf/aaai/ChiarielloMP22}, through the same automaton encoding we have been referring to throughout the paper, and instead an exhaustive search-based, \declare-specific implementation is provided in the \textsf{Declare4Py}~\cite{DBLP:conf/bpm/DonadelloRMS22} library. From the ASP perspective, a conformance checking encoding can be easily adapted to perform query checking, by searching over possible variable-activities bindings that yield a constraint above the chosen support threshold. In particular, we adapt the query checking encoding presented in~\cite{DBLP:conf/kr/FiondaIR23} to the \LTLf setting. In order to encode the query checking problem, we slightly change our input model representation, as reported in the following example.

\begin{example}[Query checking] Consider the query checking problem instance over the template $Response$, with both its activation and target ranging over $\mathcal{A}$. The \texttt{var\_bind/3} predicate, analogously to \texttt{bind\_3}, models that in a given template a parameter is bound to a variable. For the query checking problem, we are interested in tuples of activities that, when substituted to the constraints' variables, yield a constraint whose support is above the threshold over the input log. The \texttt{domain/2} predicate can be used to give each variable its own subset of possible values, but in this case, for both variables, the domain of admissible substitutions spans over $\mathcal{A}$. The choice rule generates candidate substitutions that are pruned by the constraints if they are above the maximum number of violations. Given an input support threshold $s \in (0, 1]$, the constant \texttt{max\_violations} is set to the nearest integer above $(1 - s) \cdot |\mathcal{L}|$.
\begin{lstlisting}
constraint(c,"Response"). var_bind(c,arg_0,var(a)). var_bind(c,arg_1,var(b)).
domain(var(a),A) :- trace(_,_,A).
domain(var(b),A) :- trace(_,_,A).
{ bind(C,Arg,Value): domain(Var,Value) } = 1 :- var_bind(C,Arg,Var).
:- #count{X: not sat(C, X), constraint(C,_), trace(X, _, _)} > max_violations.
\end{lstlisting}
{Notice the ASP formulation can be easily generalized (by simply adding facts encoding a \declare constraint and slightly modifying the final constraint) to query check entire \declare models (e.g., multiple constraints, where a variable might occur as activation/target of distinct constraints), rather than a single constraint at a time, by adding the following facts:}

\begin{lstlisting}
constraint(c,"Precedence"). var_bind(c,arg_0,var(a)). var_bind(c,arg_1,var(c)).
domain(var(c),A) :- trace(_,_,A).
\end{lstlisting}

Here, the variable $a$ is the activation of the $Response$ constraint (as before), as well as the target of the $Precedence$ constraint.
\end{example}

\section{Direct ASP encoding for \declare} The encodings described in previous section are general techniques that enable reasoning over arbitrary \LTLp formulae. {Both encodings require, respectively, to keep track of each subformula evaluation on each time-point of a trace in the case of the syntax-tree encoding, and to keep track of DFA state during the trace traversal, in the case of the automaton encoding. However, \declare patterns do not involve \textit{complex} temporal reasoning, with deep nesting of temporal operators. Hence \declare templates admit a succinct and direct encoding in ASP that does not keep track of such evaluation at each time-point of the trace. The encoding discussed in this section exploits this, providing an ad-hoc, direct translation of the semantics of \declare constraints into ASP rules.} 

The general approach we follow in defining the templates, is to model cases in which constraints fail through a \texttt{fail/2} predicate. {In our encoding, a constraint $c$ over a trace $tid$ holds true if we are unable to produce the atom $fail(c,tid)$.} Due to the activation-target semantics of \declare templates, sometimes it is required to assert that an activation condition is matched in the suffix of the trace by a target condition. In the encoding, this is modeled by the \texttt{witness/3} predicate. This mirrors the \emph{activation} and \emph{target} concepts in the definition of \declare constraints. Note that, this approach is not based on a systematic, algorithmic rewriting, {but on a template-by-template ad-hoc analysis. In the rest of the section, we show the principles behind our ASP encoding for the \declare templates in Table~\ref{tab:declare_templates}.}

\myparagraph{Response template}
{Recall from Table~\ref{tab:declare_templates} that $\text{Response}(a,b)$ is defined as the \LTLp formula $\Always(a \myimplies \Eventually b)$, whose informal meaning is that \emph{whenever $a$ happens, $b$ must happen somewhere in the future}. Thus, every time we observe an $a$ at time $t$, in order for $Response(a,b)$ to be true, we have to observe $b$ at a time instant $t' \geq t$. The first rule below encodes this situation. If we observe at least one $a$ that is not matched by any $b$ in the future, the constraint fails, as encoded in the second rule. {The following equivalences exploit the duality of temporal operators $\Always$, and $\Eventually$:}}
{$$\lnot \Always(a \rightarrow \Eventually b) \equiv 
\Eventually \lnot (a \rightarrow \Eventually b) \equiv \Eventually \lnot (\lnot a \lor \Eventually b) \equiv \Eventually (a \land \lnot \Eventually b)$$}

{Thus, we encode $Response$ failure conditions with the following logic program:}

\begin{minipage}{\linewidth}
\centering
\begin{minipage}{0.45\linewidth}
\begin{lstlisting}
witness(C,T,TID) :- 
  constraint(C, "Response"), 
  bind(C,arg_0,X), 
  bind(C,arg_1,Y), 
  trace(TID,T,X),
  trace(TID,T',Y), T'>=T.


\end{lstlisting}
\end{minipage}
\hspace{0.08\linewidth}
\begin{minipage}{0.45\linewidth}
\begin{lstlisting}
fail(C,TID) :-
  constraint(C,"Response"), 
  bind(C,arg_0,X), 
  bind(C,arg_1,Y),
  trace(TID,T,X), 
  not witness(C,T,TID).
\end{lstlisting}
\end{minipage}
\end{minipage}

{The rule above on the left yields a $witness(c,t,tid)$ atom whenever trace $tid$ at time $t$ satisfies $a \land \Eventually b$. In the rule on the right, the $fail(c,tid)$ atom models that there exists some time-point $t$ such that $\pi_t = a$, but $\lnot (a \land \Eventually b)$ thus (since $a \in \pi_t$) that $\lnot \Eventually b$; that is, there exists a time-point $t$ such that $a \land \lnot \Eventually b$.%
} {Subsection~\ref{appendix:response_picture} in the appendix exemplifies this argument graphically.}

\myparagraph{Precedence template}
{Recall from Table \ref{tab:declare_templates} that the constraint $\text{Precedence}(x,y)$ is defined as the \LTLp formula $\lnot y \WUntil x = \Always(\lnot y) \ \lor \  \lnot y \Until x$, whose informal meaning is that \emph{if $y$ occurs in the trace, $x$ must have happened before}. Notice that in order to witness the failure of this constraint, it is enough to reason about the trace prefix up to the first occurrence of $y$, since in the definition of the template the until operator is not under the scope of a temporal operator. {The following equivalences hold:}
$$\lnot( \lnot y \WUntil x ) \equiv \lnot (\Always(\lnot y) \lor \lnot y \Until x) \equiv \lnot \Always(\lnot y) \land \lnot (\lnot y \Until x) \equiv \Eventually y \land \lnot (\lnot y \Until x) \equiv \Eventually y \land y \Release \lnot x$$}

{We model this failure condition with the following logic rules:}

\begin{minipage}{\linewidth}
\centering
\begin{minipage}{0.45\linewidth}
\begin{lstlisting}
fail(C,TID) :-
  constraint(C, "Precedence"),
  bind(C, arg_0, X), 
  bind(C, arg_1, Y), 
  trace(TID,T',Y),
  T = #min{Q: trace(TID,Q,X)},
  trace(TID,T,X),
  T' < T.
\end{lstlisting}
\end{minipage}
\hspace{0.08\linewidth}
\begin{minipage}{0.45\linewidth}
\begin{lstlisting}
fail(C,TID) :-
  constraint(C, "Precedence"),
  bind(C, arg_0, X), 
  bind(C, arg_1, Y), 
  trace(TID,_,Y),
  not trace(TID,_,X).
\end{lstlisting}
\end{minipage}
\end{minipage}
{The rule on the left models the formula $(y \Release \lnot x)$, that is satisfied by traces where $x$ does not occur up to the point where $y$ first becomes true. The rule on the right models traces where $x$ does not occur at all, but $y$ does. We model this separately to be compliant with \texttt{clingo}'s behavior where \texttt{\#min} would yield the special term \texttt{\#sup} over an empty set of literals.} {Subsection~\ref{appendix:precedence_picture} in the appendix exemplifies how the rules model the constraint graphically.}

\medskip

{Next we consider the $AlternateResponse$ and $AlternatePrecedence$ templates. Differently from the previous cases, mapping out failure from the \LTLp formula is less intuitive, due to temporal operator nesting in the template definitions, and requires more care.}

\myparagraph{AlternateResponse template}
{Recall from Table~\ref{tab:declare_templates} that $AlternateResponse(a,b)$ is defined as the \LTLp formula $\Always(a \rightarrow \Next(\lnot a \Until b))$, whose informal meaning is that \emph{whenever $a$ happens, $b$ must happen somewhere in the future and, up to that point, $a$ must not happen.} Failure conditions follow from the this chain of equivalences:}
{$$\lnot \Always(a \rightarrow \Next(\lnot a \Until b) \equiv \Eventually (a \land \lnot \Next(\lnot a \Until b)) \equiv \Eventually(a \land \WNext \lnot(\lnot a \Until b)) \equiv \Eventually(a \land \WNext (a \Release \lnot b))$$}

{Thus, the failure condition is to observe an $a$ at time $t$, such that $\WNext(a \Release \lnot b)$ holds, that is an occurrence of $a$ at time $t' > t$ with $b \not\in \pi_k$ for all $t < k < t'$. To model the constraint, we ensure at least one $b$ appears in-between the $a$ occurrences. We model this by the following rules:}

\begin{minipage}{\linewidth}
\centering
\begin{minipage}{0.45\linewidth}
\begin{lstlisting}
witness(C,T,TID) :-
  constraint(C, "Alternate Response"),
  bind(C, arg_0, X), 
  bind(C, arg_1, Y), 
  trace(TID,T,X),
  T'' = #min{Q: 
    trace(TID,Q,X), Q > T
  }, 
  trace(TID,T',Y), 
  T'' > T', 
  T' > T.

\end{lstlisting}
\end{minipage}
\hspace{0.08\linewidth}
\begin{minipage}{0.45\linewidth}
\begin{lstlisting}
fail(C,TID) :- 
  constraint(C, "Alternate Response"),
  bind(C, arg_0, X), 
  trace(TID,T,X), 
  not witness(C,T,TID).
  
\end{lstlisting}
\end{minipage}
\end{minipage}
{A witness here is observing an $a$  at time $t$ such that the next occurrence of $a$ at time $t'$, with at least one $b$ in-between. To get this succinct encoding, we rely on \texttt{clingo} \texttt{\#min} behavior: \texttt{\#min} of an empty term set is the constant \texttt{\#sup}, so the arithmetic literal \texttt{T'' > T'} will always be true when there's no occurrence of $a$ following the one at time $T$ - this deals with the ``last'' occurence of $a$ in the trace. If two ``adjacent'' occurrences of $a$  do not have a $b$ in-between, that does not count as a witness (as by the constraint's semantics). 
{Subsection~\ref{appendix:alternate_response_picture} in the appendix exemplifies this argument graphically.}

\myparagraph{AlternatePrecedence template} {Recall from Table \ref{tab:declare_templates} that the constraint $AlternatePrecedence(x,y)$ is defined as the \LTLp formula 
{$$(\lnot y \WUntil x) \land \Always(b \rightarrow \WNext (\lnot y \WUntil x)) = Precedence(a,b) \land \Always(b \rightarrow \WNext Precdence(a,b))$$}
\noindent whose informal meaning is that \emph{every time $y$ occurs in the trace, it has been preceded by $x$ and no $y$ happens in-between}. To map its failure conditions, we build this chain of equivalences, where $\alpha = Precedence(a,b)$:}
{$$\lnot (\alpha \land \Always(y \rightarrow \WNext \alpha)) \equiv \lnot \alpha \lor \Eventually(y \land \lnot \WNext \alpha)$$}
{Above we described the encoding of $Precedence(x,y)$, thus, now we focus on the second term:}
{$$\Eventually(y \land \lnot \WNext \alpha) \equiv \Eventually(y \land \lnot last \land \lnot \Next \alpha) \equiv \Eventually(y \land \lnot last \land \WNext \lnot\alpha) \equiv \Eventually(y \land \Next \lnot\alpha)$$}
where $last$ is a propositional symbol that is true only in the last state of the trace (that is, $last \equiv \WNext \lnot\top$). {Finally, by substituting $\lnot \alpha$ with the failure condition of $Precedence(x,y)$}:
{$$\Eventually(y \land \Next \lnot \alpha) \equiv \Eventually(y \land \Next(\Eventually y \land y \Release \lnot x))$$}
{Thus, $AlternatePrecedence(x,y)$ admits the same failure conditions as $Precedence(a,b)$, that are reported in the rules on the left below. Furthermore, from the second term we derive the failure condition which regards  the occurrence of a $b$ that is followed by a trace suffix where $Precedence(a,b)$ does not hold; that is, the $b$ is followed by another $b$ with no $a$ in between. This is modeled by the rule on the right:}

\begin{minipage}{\linewidth}
\centering
\begin{minipage}{0.45\linewidth}
\begin{lstlisting}
fail(C,TID) :-
  constraint(C, "Alternate Precedence"),
  bind(C, arg_0, X), 
  bind(C, arg_1, Y), 
  trace(TID,T',Y), 
  T = #min{Q: trace(TID,Q,X)},
  trace(TID,T,X),  
  T' < T.

fail(C,TID) :-
  constraint(C, "Alternate Precedence"),
  bind(C, arg_0, X), 
  bind(C, arg_1, Y), 
  trace(TID,_,Y), 
  not trace(TID,_,X).
\end{lstlisting}
\end{minipage}
\hspace{0.08\linewidth}
\begin{minipage}{0.45\linewidth}
\begin{lstlisting}

fail(C,TID) :-
  constraint(C, "Alternate Precedence"), 
  bind(C, arg_0, X), 
  bind(C, arg_1, Y), 
  trace(TID, T0, Y), 
  trace(TID, T2, Y),
  #count{Q: 
    trace(TID,Q,X), 
    Q >= T0, Q <= T2
  } = 0,
  T2 > T0.

\end{lstlisting}
\end{minipage}
\end{minipage}

{Subsection~\ref{appendix:alternate_precedence_picture} in the appendix exemplifies this argument graphically.}

\myparagraph{ChainResponse template} {Recall that from Table~\ref{tab:declare_templates} that the constraint $ChainResponse(x,y)$ is defined as the \LTLp formula $\Always(x \rightarrow \Next y)$, whose informal meaning is that \emph{whenever $x$ occurs, it must be immediately followed by $y$}. The failure conditions follow from the following chain of equivalences:}
{$$\lnot(\Always(x \rightarrow \Next y)) \equiv \Eventually \lnot (x \rightarrow \Next y) \equiv \Eventually \lnot(\lnot x \lor \Next y) \equiv \Eventually(x \land \lnot \Next y) \equiv \Eventually(x \land \WNext \lnot y)$$}

{That is, an $x$ occurs that is not followed by a $y$ - either because $x$ occurs as the last activity of the trace or that $x$ is followed by a different activity. We can model this by the following rule:}

\begin{lstlisting}
fail(C,TID) :- 
  constraint(C, "Chain Response"),
  bind(C, arg_0, X), bind(C, arg_1, Y), 
  trace(TID,T,X), not trace(TID,T+1,Y).
\end{lstlisting}

{Subsection~\ref{appendix:chain_response_picture} in the appendix exemplifies this argument graphically.}

\myparagraph{ChainPrecedence template} {Recall that from Table~\ref{tab:declare_templates} that the constraint $ChainResponse(x,y)$ is defined as the \LTLp formula $\Always(\Next y \rightarrow x) \land \lnot b$, whose informal meaning is that \emph{whenver $y$ occurs, it must have been immediately preceded by $x$}. The failure conditions follow from the following chain of equivalences:}
{$$\lnot(\Always(\Next y \rightarrow x) \land \lnot y) \equiv (\Eventually \lnot (\Next y \rightarrow x) \lor y \equiv \Eventually(\Next y \land \lnot x) \lor y$$}

{In this case, $ChainPrecedence$ fails if the predecessor of $y$ is not $x$, or if $y$ occurs in the first instant of the trace. This is modeled by the following rules:}

\begin{lstlisting}
fail(C, TID) :- 
  constraint(C, "Chain Precedence"),
  bind(C, arg_0, X), bind(C, arg_1, Y),
  trace(TID,T+1,Y), trace(TID,T,_),     
  not trace(TID,T,X).
fail(C, TID) :-
  constraint(C, "Chain Precedence"), 
  bind(C, arg_1, Y), trace(TID, 0, Y).
\end{lstlisting}

{Subsection~\ref{appendix:chain_precedence_picture} in the appendix exemplifies this argument graphically.}

{\myparagraph{Modeling the Succession hierarchy} Templates in the $Succession$ chain are defined as the conjunction of the respective $Precedence, Response$ templates at the same ``level of the subsumption hierarchy (see Figure~\ref{fig:subsumption_hierarchy}); thus, it is possible to encode them in the \texttt{fail/2}, \texttt{witness/3} schema, since $\lnot(\varphi \land \psi) = \lnot \varphi \lor \lnot \psi$. Hence, the failure conditions for $Succession$-based templates are the union of the failure conditions of its sub-formulae (all the cases were thoroughly described above in this section).}

{\myparagraph{Modeling non-temporal templates} 
Choice (i.e., $Choice$, $ExclusiveChoice$) and Existential templates (i.e., $RespondedExistence$, $Coexistence$) are defined only in terms of conjunction and disjunction of atomic formulae (e.g., activity occurrences in the trace). Thus, they are easy to implement using normal rules whose body contains \texttt{trace/3} literals to model that $a \in \pi_i, a \not\in \pi_i$. As an example, consider the constraint $RespondedExistence(a,b)$ that states that \emph{whenever $a$ occurs in the trace, $b$ must occur as well}}. {Its \LTLp definition is $\Eventually(a) \rightarrow \Eventually(b)$, its failure conditions follow from the following chain of equivalences:}

{$$\lnot(\Eventually(a) \rightarrow \Eventually(b)) \equiv \lnot(\lnot \Eventually(a) \lor \Eventually(b)\equiv \Eventually(a) \land \lnot \Eventually(b)$$}

\begin{lstlisting}
fail(C,TID) :- constraint(C, "Responded Existence"),
  bind(C, arg_0, X), bind(C, arg_1, Y), trace(TID,_,X), not trace(TID,_,Y).
\end{lstlisting}

{All the encodings for the \declare constraints in Table~\ref{tab:declare_templates} are available in the repository%
\footnote{\repositoryurl.}.}

\begin{table}[b!]
    \centering 
    \caption{Log Statistics: $|\mathcal{A}|$ is the number of activities; Average $|\pi|$ is the average trace length; $|\mathcal{L}|$ is the number of traces; $|\mathcal{C}^{\text{IV}}|$ is the number of \declare constraints above 50\% support.}
    \label{tab:log_stats}
    \begin{tabular} {r@{\hspace{1em}}r@{\hspace{1em}}r@{\hspace{1em}}r@{\hspace{1em}}r@{\hspace{1em}}r@{\hspace{1em}}r}
        \hline
        \textbf{Log name} & {$|\mathcal{A}|$} & {Average $|\pi|$} & {Max $|\pi|$} & {Min $|\pi|$} & {$|\mathcal{L}|$} & $|\mathcal{C}^{\text{IV}}|$ \\
        \hline
        Sepsis Cases (SC) & 16 & 14.5 & 185 & 3 & 1050 & 76\\
        Permit Log (PL) & 51 & 12.3 & 90 & 3 &7065 & 26\\
        BPI Challenge 2012 (BC)  & 23 & 12.6 & 96 & 3 & 13087 & 10\\
        Prepaid Travel Cost (PC) & 29 & 8.7 & 21 & 1 &2099 & 52\\
        Request For Payment (RP) & 19 & 5.4 & 20 & 1 &6886 & 52\\
        International Declarations (ID) & 34 & 11.2 & 27 & 3 &6449 & 152\\
        Domestic Declarations (DD) & 17 & 5.4 & 24 & 1 &10500 & 52\\
        \hline
    \end{tabular}
\end{table}

\begin{center}
\begin{figure}[t]
    \begin{minipage}{0.45\linewidth}
    \hspace{-0.1\linewidth}
    \begin{tikzpicture}[scale=0.70]
            \begin{axis}[
                title={},
                xlabel={Solved conformance checking instances},
                ylabel={Execution time (s)},
                ymode=log,
                legend pos=north west
            ]
            \addplot[
                color=magenta,
                mark=x,
                ]
                table {plots/nc/cf_asp_native.data};
                \addlegendentry{$\text{ASP}_\mathcal{D}$}
            \addplot[
                color=blue,
                mark=diamond,
                ]
                table {plots/nc/cf_d4py.data};
                \addlegendentry{\textsf{D4Py}}
            \addplot[
                color=cyan,
                mark=pentagon,
                ]
                table {plots/nc/cf_automata.data};
                \addlegendentry{$\text{ASP}_\mathcal{A}$}
            \addplot[
                color=orange,
                mark=square,
                ]
                table {plots/nc/cf_ltlf_base.data};
                \addlegendentry{$\text{ASP}_\mathcal{S}$}
            \end{axis}
            \end{tikzpicture}
            \caption{Conformance checking cactus.}
            \label{fig:cf_plot}
    \end{minipage}
    \begin{minipage}{0.45\linewidth}
    \begin{table}[H]
    {\small
        \resizebox{\textwidth}{!}{
        \begin{tabular}{r@{\hspace{1em}}r@{\hspace{1em}}r@{\hspace{1em}}r@{\hspace{1em}}r}
            \hline
            \textbf{Log} & $\text{ASP}_{\mathcal{D}}$ & \textsf{D4Py} & $\text{ASP}_{\mathcal{A}}$ & $\text{ASP}_{\mathcal{S}}$ \\
            \hline
        ID & \textbf{23.3} & 39.5 & 124.9 & 4621.7\\
        RP & \textbf{10.8} & 16.3 & 25.8 & 409.8\\
        PT & \textbf{5.2} & 8.5 & 12.6 & 121.6\\
        SC & \textbf{4.3} & 11.6 & 13.4 & 141.2\\
        PL & \textbf{10.8} & 35.6 & 20.7 & 624.1\\
        DD & \textbf{14.2} & 22.4 & 40.1 & 963.2\\
        BC & \textbf{14.1} & 23.7 & 20.5 & 796.6\\
        \hline
        \end{tabular}}
        \caption{Runtime in seconds for conformance checking on $\mathcal{C}^{\text{IV}}$.}
        \label{tab:cf_table}
        }
    \end{table}
    \end{minipage}
\end{figure}
\end{center}

\section{Experiments}
In this section, we report the results of our experiments comparing different methods to perform conformance checking and query checking of \declare models, using the ASP-based representations outlined in the previous sections and \textsf{Declare4Py}, {a recent Python library for \declare-based process mining tasks which also implements - among other tasks, such as \textit{log generation} and \textit{process discovery} - conformance checking and query checking functionalities. While the approaches discussed here are declarative in nature, \textsf{Declare4Py} implements \declare semantics by imperative procedures, based on the algorithms in ~\cite{DBLP:journals/eswa/BurattinMS16}, that scan the input traces. \textsf{Declare4Py}  supports also other tasks, and most importantly, supports a data-aware variant of \declare, {that take into account data attributes associated to events in a trace}, while here we deal with standard, control-flow only \declare.} Methods will be referred to as $\text{ASP}_\mathcal{D}$, $\text{ASP}_\mathcal{A}$, $\text{ASP}_\mathcal{S}$ and \textsf{D4Py} --- denoting respectively our direct encoding, the automata and syntax tree-based translation methods and \textsf{Declare4Py}. We start by describing datasets (logs and \declare models), and execution environment to conclude by discussing experimental results. Subsection~\ref{exp:tracescaling} encompass further experimental analysis about memory consumption and behavior on longer traces for our ASP encoding.

\myparagraph{Data} %
We validate our approach on real-life event logs from past BPI Challenges~\cite{DBLP:conf/bpm/LopesF19}. These event logs are well-known and actively used in Process Mining literature.
For each event log $\mathcal{L}_i$, we use \dpy to mine the set of \declare constraints $\mathcal{C}_i$ whose support on $\mathcal{L}_i$ is above 50\%. Then, we define four models, $\mathcal{C}^{\text{I}}_i, \mathcal{C}^{\text{II}}_i, \mathcal{C}^{\text{III}}_i, \mathcal{C}^{\text{IV}}_i$, containing respectively the first 25\%, 50\%, 75\% and 100\% of the constraints in a random shuffling of $\mathcal{C}_i$, such that $\mathcal{C}^{\text{I}}_i \subset \mathcal{C}^{\text{II}}_i \subset \mathcal{C}^{\text{III}}_i \subset  \mathcal{C}^{\text{IV}}_i$. Table~\ref{tab:log_stats} summarizes some statistics about the logs and the \declare models we mined over the logs. All resource measurements take into account the fact that ASP encodings require an additional translation step from the XML-based format of event logs to a set of facts. The translation time is included in the measurement times and is comparable with the time taken by \dpy. %

\myparagraph{Execution Environment} The experiments in this section were executed on an Intel(R) Xeon(R) Gold 5118 CPU @ 2.30GHz, 512GB RAM machine, using \textsc{Clingo} version 5.4.0, Python 3.10, \dpy 1.0 and \texttt{pyrunlim}\footnote{\href{https://github.com/alviano/python/tree/master/pyrunlim}{https://github.com/alviano/python/tree/master/pyrunlim}} to measure resources usage. Experiments were run sequentially. All data and scripts to reproduce our experiments are available in \href{\repository}{the repository}.

\subsection{Conformance checking \& Query checking on real-world logs}
\begin{center}
\begin{figure}[t]
    \begin{minipage}{0.45\linewidth}
    \hspace{-0.1\linewidth}
\begin{tikzpicture}[scale=0.70]
    \begin{axis}[
        title={},
        xlabel={Solved query checking instances},
        ylabel={Execution time (s)},
        ymode=log,
        legend pos=north west
    ]
    \addplot[
        color=magenta,
        mark=x,
        ]
        table {plots/nc/qc_asp_native.data};
                \addlegendentry{$\text{ASP}_\mathcal{D}$}
    \addplot[
        color=blue,
        mark=diamond,
        ]
        table {plots/nc/qc_d4py.data};
        \addlegendentry{\textsf{D4Py}}
    \addplot[
        color=cyan,
        mark=pentagon,
        ]
        table {plots/nc/qc_automata.data};
                \addlegendentry{$\text{ASP}_\mathcal{A}$}
    \addplot[
        color=orange,
        mark=square,
        ]
        table {plots/nc/qc_ltlf_base.data};
                \addlegendentry{$\text{ASP}_\mathcal{S}$}
    \end{axis}
    \end{tikzpicture}    
    \caption{Query checking cactus plot.}        \label{fig:qc_plot}
    \end{minipage}
    \begin{minipage}{0.45\linewidth}
    \begin{table}[H]
    {\small
        \resizebox{\textwidth}{!}{
        \begin{tabular}{r@{\hspace{1em}}r@{\hspace{1em}}r@{\hspace{1em}}r@{\hspace{1em}}r}
            \hline
            \textbf{Log} & $\text{ASP}_{\mathcal{D}}$ & \textsf{D4Py} & $\text{ASP}_{\mathcal{A}}$ & $\text{ASP}_{\mathcal{S}}$ \\
            \hline
        ID & \textbf{817.2} & 1624.5 & 1654.0 & 3522.4\\
        RP & 884.2 & 565.8 & \textbf{318.2} & 1179.4\\
        PT & \textbf{223.6} & 451.1 & 236.1 & 427.9\\
        SC & \textbf{163.8} & 267.0 & 173.1 & 665.1\\
        PL & \textbf{1614.0} & 4227.7 & 3926.8 & 5397.5\\
        DD & \textbf{407.7} & 698.2 & 479.2 & 2436.2\\
        BC & \textbf{2304.8} & 2467.7 & 6636.0 & 27445.3\\
        \hline
        \end{tabular}}
        \caption{Cumulative runtime in seconds for query checking tasks.}
        \label{tab:qc_table}
        }
    \end{table}
    \end{minipage}
\end{figure}
\end{center}

\myparagraph{Conformance Checking} We consider the conformance checking tasks $(\mathcal{L}_i, \mathcal{M})$, with $\mathcal{M} \in \{\mathcal{C}_i^\text{I}, \mathcal{C}_i^\text{II}, \mathcal{C}_i^\text{III}, \mathcal{C}_i^\text{IV}\}$, over the considered logs and its \declare models. Figure~\ref{fig:cf_plot} reports the solving times for each method in a cactus plot. Recall that a point $(x,y)$ in a cactus plot represents the fact that a given method solves the $x$-th instance, ordered by increasing execution times, in $y$ seconds. Table~\ref{tab:cf_table} reports the same data aggregated by the event log dimension, best run-time in bold. Overall, our direct encoding approach is faster than the other ASP-based encodings as well as \dpy on considered tasks. $\text{ASP}_\mathcal{A}$ and \dpy perform similarly, whereas $\text{ASP}_\mathcal{S}$ is less efficient. {It must also be noticed that \dpy, beyond computing whether a trace is compliant or not with a given constraint, also stores additional information such as the number of times a constraint is violated, or activate, while the ASP encodings do not. However, the automata and direct encoding can be straightforwardly extended in such sense; in particular, similar ``book-keeping'' in the direct encoding is performed by the \texttt{fail/3} and \texttt{witness/3} atoms.}

\myparagraph{Query Checking} We consider the query checking instances $(t, \mathcal{L}_i, s)$ where $t$ is a \declare template, from the ones defined in Table~\ref{tab:declare_templates}, $s \in \{0.50, 0.75, 1.00\}$ is a support threshold, and $\mathcal{L}_i$ is a log.  Figure~\ref{fig:qc_plot} summarizes the results in a cactus plot, and Table~\ref{tab:qc_table} aggregates the same data on the log dimension, best runtime in bold. $\text{ASP}_\mathcal{D}$ is again the best method overall, outperforming other ASP-based methods with the exception of $\text{ASP}_\mathcal{A}$ on the RP log tasks. Again, $\text{ASP}_\mathcal{A}$ and \textsf{D4Py} perform similarly and $\text{ASP}_\mathcal{S}$ is the worst. 

\myparagraph{Discussion} In the comparative evaluation of the different methods for conformance and query checking tasks, our direct encoding approach $\text{ASP}_\mathcal{D}$ showed better performance compared to both \textsf{D4Py} and other ASP-based methods, as reported in Table~\ref{tab:cf_table} and Table~\ref{tab:qc_table} (and in Fig.~\ref{fig:cf_plot} and Fig.~\ref{fig:qc_plot}). As shown in our experiments, $\text{ASP}_\mathcal{D}$ not only outperformed in terms of runtime the other methods but also offers a valuable alternative when considering overall efficiency. Notably, $\text{ASP}_\mathcal{A}$ and \textsf{D4Py} showed similar performances, with $\text{ASP}_\mathcal{A}$ slightly outperforming in certain instances, particularly in the RP log, but $\text{ASP}_\mathcal{S}$ exhibited less efficiency in nearly all the logs.

From the point of view of memory consumption (Table~\ref{tab:maxmem}), \textsf{D4Py} proved to be the most efficient in query checking tasks. This can be attributed to its imperative implementation that allows for an \emph{``iterate and discard''} approach to candidate assignments, avoiding the need for their explicit grounding required by ASP-based techniques. $\text{ASP}_\mathcal{S}$ is the least efficient method regarding memory usage, consistently across both types of tasks and all logs. We conjecture the significant increase in maximum memory usage is the primary factor contributing to the runtime performance degradation in both tasks, {that might make the encodings an interesting benchmark for compilation-based ASP systems~\cite{DBLP:conf/aaai/MazzottaRD22,DBLP:conf/kr/DodaroMR24,DBLP:conf/ecai/DodaroMR23,DBLP:conf/cilc/CuteriMR23}}. In fact, $\text{ASP}_\mathcal{S}$ proved to be the less efficient in both memory consumption and running time. Furthermore, we observe that in conformance checking $\text{ASP}_\mathcal{D}$ is more efficient w.r.t. memory consumption when compared to \textsf{D4Py}, and, when combined, these two methods collectively show better memory efficiency when compared to other ASP-based methods, translating in lower running times.

The obtained results clearly indicates the lower memory requirements of \textsf{D4Py}, demonstrating its applicability in resource-constrained environments. However, the compact and declarative nature of ASP provides an efficient means to implement \declare constraints, as demonstrated by the performance of $\text{ASP}_\mathcal{D}$, which is especially suitable in environments where memory is less of a constraint and execution speed is a key factor. Furthermore, the ASP-based approaches can be more readily extended, in a pure declarative way, to perform {, for example, } query-checking of multiple \declare constraints in a matter of few extra rules.

\begin{table}[t]        
    \resizebox{\textwidth}{!}{
    \begin{tabular}{r@{\hspace{1em}}r@{\hspace{1em}}r@{\hspace{1em}}r@{\hspace{1em}}r@{\hspace{1em}}r@{\hspace{1em}}r@{\hspace{1em}}r@{\hspace{1em}}r}
    \hline
    \multicolumn{1}{c}{\multirow{2}{*}{\textbf{Log}}} & \multicolumn{4}{c}{\textbf{Conformance Checking}} & \multicolumn{4}{c}{\textbf{Query Checking}} \\
         & $\text{ASP}_{\mathcal{D}}$ & \textsf{D4Py} & $\text{ASP}_{\mathcal{A}}$ & $\text{ASP}_{\mathcal{S}}$ & $\text{ASP}_{\mathcal{D}}$ & \textsf{D4Py} & $\text{ASP}_{\mathcal{A}}$ & $\text{ASP}_{\mathcal{S}}$ \\
        \hline
        BC & \textbf{323.6} & 566.3 & 546.0 & 8757.9 & 3157.0 & \textbf{580.7} & 1450.8 & 18866.5\\
        DD & 536.6 & \textbf{386.0} & 927.2 & 14473.8 & 728.1 & \textbf{336.4} & 513.0 & 2706.0\\
        ID & 837.1 & \textbf{578.4} & 2338.9 & 55435.2 & 1498.5 & \textbf{583.5} & 763.6 & 5029.8\\
        PL & \textbf{312.8} & 2062.2 & 579.5 & 11388.3 & 2434.2 & 2071.0 & \textbf{1023.3} & 7006.2\\
        PC & \textbf{222.2} & 281.9 & 341.2 & 5211.2 & 435.3 & 283.4 & \textbf{282.8} & 1209.3\\
        RP & 372.3 & \textbf{347.6} & 610.8 & 9706.0 & 574.8 & \textbf{312.3} & 396.9 & 1843.3\\
        SC & \textbf{195.0} & 245.6 & 336.4 & 5786.5 & 480.4 & \textbf{244.6} & 247.2 & 2146.7\\
    \hline
    \end{tabular}}
    \caption{Max memory usage (MB) over all the conformance checking and query checking tasks, aggregated by log, for all the considered methods. Lowest value in boldface.}
    \label{tab:maxmem}
\end{table}

{An intuitive reason for such a memory usage gap in the conformance checking tasks is the following. As noted in the encodings section, both the automata encoding and the syntax tree encoding rely on a 4-ary predicate ($true(TID,C,X,T), cur\_state(TID,C,X,T)$ respectively), that keeps track of subformulae evaluation on each instant of the trace and current state in the constraint DFA for each instant of the trace. In the case of the automaton encoding, $X$ does not index subtrees but states of the automaton - all the other terms have the same meaning. During the grounding of the logic programs, this yields a number of symbolic atoms that scales linearly w.r.t. the total number of events in the log (that is, the sum of lenghts of the traces in the log) and linearly in the number of nodes in the syntax tree/states in the automaton. The gap between the formula encoding and the automata encoding is explained by the fact that the \LTLf to symbolic automaton translation, albeit 2-EXP in the worst case, results usually in \textit{more compact} automatons in the case of the \declare patterns. For example, considering the Response template: its formula definition is $\Always(a \rightarrow \Next b)$, with a size (number of subformulae) of 5, while its automaton as 2 states. Furthermore, since \declare assumes simplciity, (e.g., has $|\pi_i| = 1$), the symbolic automaton can be further simplified for some constraints. This yields, overall, \textit{less states w.r.t the number of nodes in the parse tree} - that reduces the number of ground symbols. Total number of events is the same for both encodings, since both encodings share the same input fact schema. On the other hand, as we sketched in Section 3, the direct encoding explicitly models \textit{how constraints are violated}, thus for most templates, it produces less atoms, since we yield at most one \texttt{witness/3} atom for each occurrence of the constraint activation, and at most one \texttt{fail/2} atom for each constraint and each trace. Table~\ref{tab:grounding_impact} reports the number of generated symbols on the three different encodings to conformance check $\mathcal{C}^{\text{IV}}$ on the Sepsis Cases event log, where (see Table~\ref{tab:cf_table}) conformance checking shows a noticeable wide gap in runtime between the encodings, although remaining manageable runtime-wise for all the three ASP encodings. This confirms our intuitive explanation. In the case of the query checking task, being a classic guess \& check ASP encoding, the performance depends on many factors, and it is difficult to pinpoint - size of the search space (quadratic in $|\mathcal{A}|$), order of branching while searching the model, number of constraints grounded, the nature itself of traces in the log. In this context, trade-offs between runtime and memory consumption are expected.}

\begin{table}[t]
\centering
\begin{tabular}{lccc}
\hline
\multirow{2}{*}{\textbf{Metric}} & \multicolumn{3}{c}{\textbf{Method}} \\
 & $\text{ASP}_{\mathcal{D}}$ & $\text{ASP}_{\mathcal{S}}$ & $\text{ASP}_{\mathcal{A}}$ \\
\hline
Symbols & 117176 & 4593770 & 1329322 \\
Rules & 143106 & 5953435 & 1345738 \\
Execution time (s) & 0.59 & 123.55 & 10.75 \\
Max Memory (MB) &  60.94 & 5756.16 & 300.81 \\
\hline
\end{tabular}
\caption{Metrics comparison for conformance checking of $\mathcal{C}^{\text{IV}}$ over the Sepsis Cases Event Log, with ASP encodings $\text{ASP}_{\mathcal{D}}$, $\text{ASP}_{\mathcal{S}}$, and $\text{ASP}_{\mathcal{A}}$. Execution times do not take into account XES input parsing and output parsing (as in Table~\ref{tab:cf_table} , but are performed directly on facts representation of the input. Thus, reported times slightly differ from Table~\ref{tab:cf_table}, although being executed on the same log.}
\label{tab:grounding_impact}
\end{table}

\subsection{Behavior on longer traces}\label{exp:tracescaling}
{Lastly, we analyze how the automata-based and direct ASP encodings scale w.r.t the length of the traces. We generate synthetic event logs (details in the appendix) of 1000 traces of fixed length {(up to 1000 events)} for each constraint, and we perform conformance checking with respect to a single constraint, chosen among the $Response$ and $Precedence$ hierarchies. Table~\ref{tab:trace_length_table} reports the result of our experiment. We compare only automata and direct encodings, since the previous analysis make it evident the syntax tree encoding is subpar due to memory consumption. The direct encoding yields a slight, but consistent, advantage time-wise w.r.t the automata encoding, and about half of peak memory consumption. Recall this is for a \textit{single} constraint, and usually \declare models are composed of multiple \declare constraints, so this gap is expected to widen even more in conformance checking real \declare models. This is consistent with our previous experiments, see Table~\ref{tab:cf_table}.} 

{Moreover, we include a comparison between the direct encoding and \dpy over the same synthetic logs. For each synthetic log in Table~\ref{tab:trace_length_table}, a point $(x,y)$ in the scatter plot of Figure~\ref{fig:scatters} means that the conformance checking task is solved in $x$ seconds using the direct encoding and $y$ seconds using \dpy 
\footnote{Note that time measurements of Figure~\ref{fig:scatters} and Table~\ref{tab:trace_length_table} are not directly comparable, since in Table~\ref{tab:trace_length_table} input logs were stored as ASP facts, while in Figure~\ref{fig:scatters} input logs were XES files.}.
Again, these results are consistent with observed behavior over real-world event logs.}

\begin{figure}
\centering
\begin{minipage}{.7\textwidth}
  \centering
  \includegraphics[width=\linewidth]{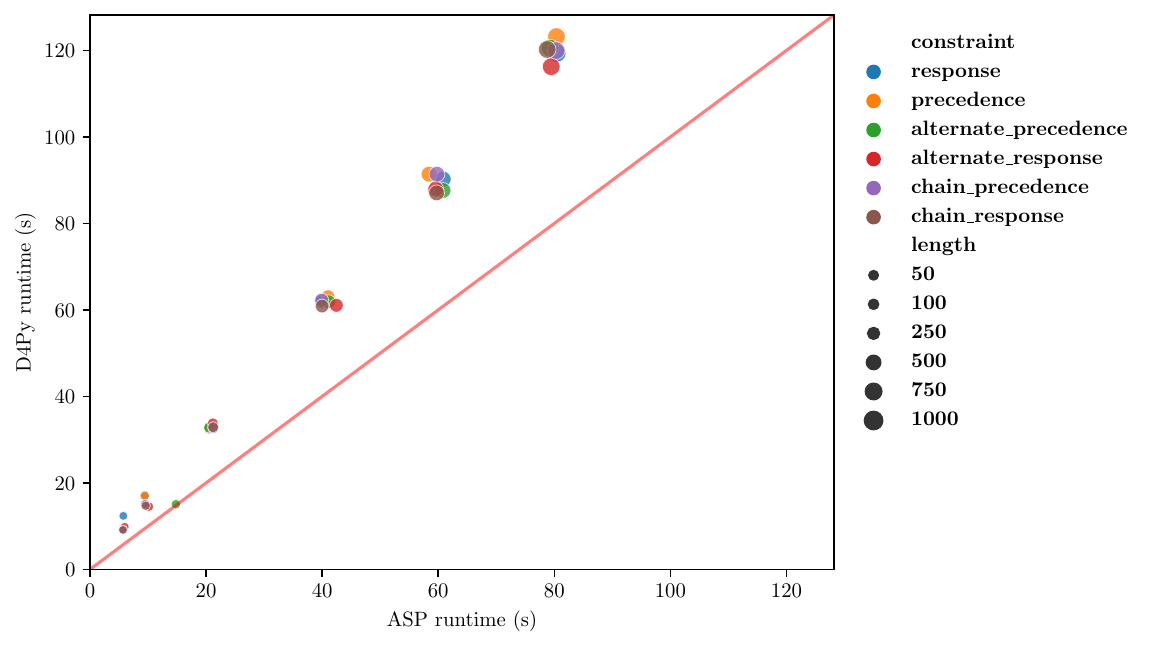}
  \label{fig:scatters_time}
\end{minipage}
\begin{minipage}{.7\textwidth}
  \centering
  \includegraphics[width=\linewidth]{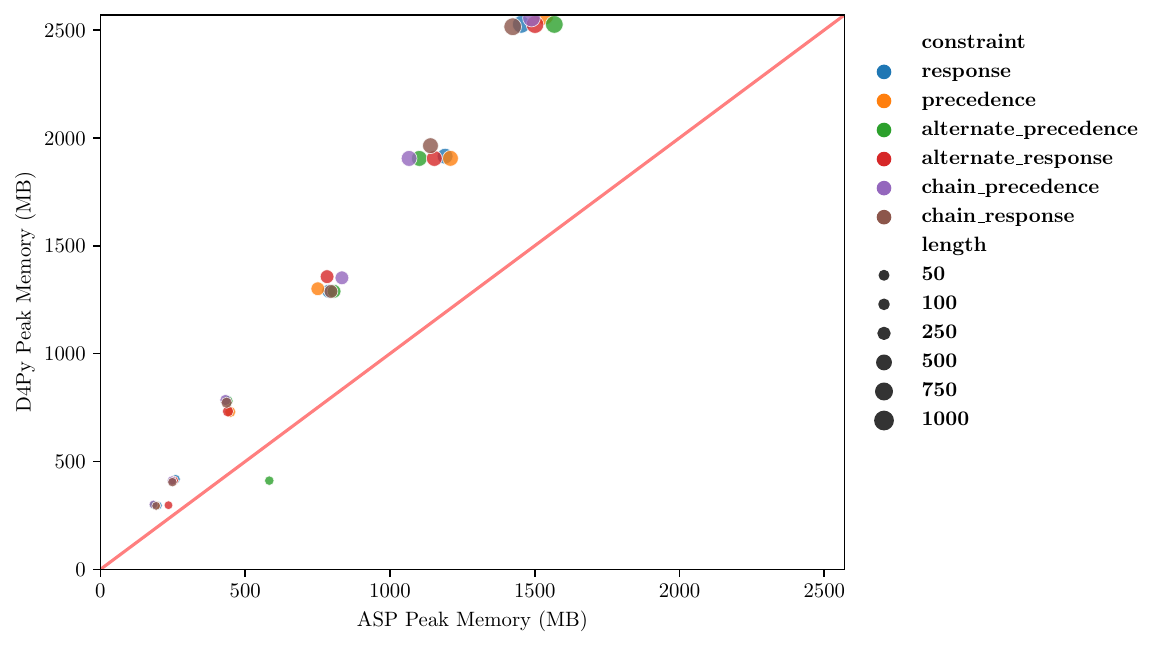}
  \label{fig:scatters_memo}
\end{minipage}%
\caption{Comparison of runtime (upper) and peak memory usage (lower) for the ASP direct encoding and \dpy, on synthetic logs used in Table~\ref{tab:trace_length_table}.}
\label{fig:scatters}
\end{figure}
\begin{table*}[t]
\centering
\resizebox{0.90\textwidth}{!}{\begin{tabular}{ccccc}
\multicolumn{3}{c}{\textbf{Response}} & \multicolumn{2}{c}{\textbf{Precedence}} \\
\multicolumn{1}{c}{$|\pi|$} & $\text{ASP}_\mathcal{D}$ & $\text{ASP}_\mathcal{A}$ & $\text{ASP}_\mathcal{D}$ & $\text{ASP}_\mathcal{A}$\\
50 & (0.898 s, 175.922 MB) & (1.603 s, 463.297 MB) & (1.055 s, 69.926 MB) & (1.186 s, 316.082 MB)\\
100 & (1.877 s, 103.254 MB) & (2.429 s, 326.863 MB) & (1.660 s, 110.941 MB) & (2.458 s, 318.984 MB)\\
250 & (4.249 s, 216.777 MB) & (7.589 s, 573.191 MB) & (4.263 s, 316.070 MB) & (6.489 s, 660.438 MB)\\
500 & (9.111 s, 401.844 MB) & (15.604 s, 871.145 MB) & (8.701 s, 385.801 MB) & (16.096 s, 845.754 MB)\\
750 & (13.121 s, 505.500 MB) & (29.809 s, 1427.172 MB) & (12.873 s, 497.191 MB) & (28.641 s, 1409.781 MB)\\
1000 & (18.693 s, 784.520 MB) & (40.228 s, 1725.414 MB) & (17.416 s, 764.488 MB) & (39.589 s, 1717.375 MB)\\
\hline

\multicolumn{3}{c}{\textbf{Alt. Response}} & \multicolumn{2}{c}{\textbf{Alt. Precedence}} \\
\multicolumn{1}{c}{$|\pi|$} & $\text{ASP}_\mathcal{D}$ & $\text{ASP}_\mathcal{A}$ & $\text{ASP}_\mathcal{D}$ & $\text{ASP}_\mathcal{A}$\\
50 & (1.397 s, 215.266 MB) & (1.679 s, 461.453 MB) & (1.016 s, 635.871 MB) & (1.751 s, 741.715 MB)\\
100 & (2.101 s, 164.242 MB) & (2.476 s, 461.355 MB) & (6.232 s, 635.867 MB) & (3.045 s, 741.715 MB)\\
250 & (4.636 s, 207.594 MB) & (7.735 s, 562.719 MB) & (4.653 s, 207.570 MB) & (8.123 s, 855.000 MB)\\
500 & (9.419 s, 404.379 MB) & (16.354 s, 869.719 MB) & (9.121 s, 401.465 MB) & (17.967 s, 995.680 MB)\\
750 & (14.057 s, 503.562 MB) & (30.529 s, 1429.836 MB) & (14.646 s, 737.258 MB) & (31.910 s, 1577.168 MB)\\
1000 & (19.415 s, 783.363 MB) & (43.013 s, 1726.945 MB) & (20.054 s, 823.027 MB) & (49.133 s, 1714.512 MB)\\

\hline
\multicolumn{3}{c}{\textbf{Chain Response}} & \multicolumn{2}{c}{\textbf{Chain Precedence}} \\
\multicolumn{1}{c}{$|\pi|$} & $\text{ASP}_\mathcal{D}$ & $\text{ASP}_\mathcal{A}$ & $\text{ASP}_\mathcal{D}$ & $\text{ASP}_\mathcal{A}$\\
50 & (1.116 s, 69.902 MB) & (1.245 s, 349.805 MB) & (0.922 s, 166.375 MB) & (1.245 s, 452.973 MB)\\
100 & (1.818 s, 349.098 MB) & (2.563 s, 587.199 MB) & (1.798 s, 166.371 MB) & (2.610 s, 452.969 MB)\\
250 & (4.972 s, 349.805 MB) & (6.891 s, 655.785 MB) & (4.678 s, 206.789 MB) & (7.862 s, 559.523 MB)\\
500 & (9.527 s, 378.543 MB) & (16.778 s, 859.781 MB) & (9.882 s, 405.285 MB) & (16.104 s, 850.094 MB)\\
750 & (13.120 s, 720.094 MB) & (27.913 s, 1539.625 MB) & (13.165 s, 744.195 MB) & (28.160 s, 1605.547 MB)\\
1000 & (17.960 s, 772.133 MB) & (40.541 s, 1731.586 MB) & (17.837 s, 794.250 MB) & (41.960 s, 1739.156 MB)\\

\hline
\end{tabular}}
\caption{Comparing metrics between different ASP encodings, with increasing trace lenghts. An entry is an ordered pair $(t, m)$ where $t$ is the runtime (seconds), $m$ the memory peak (MBs).}
\label{tab:trace_length_table}
\end{table*}

\section{Conclusion}
\declare is a declarative process modeling language, which describes processes by sets of temporal constraints. \declare specifications can be expressed as \LTLp formulae, and traditionally have been evaluated by executing the equivalent automata~\cite{DBLP:books/sp/22/CiccioM22}, regular expressions~\cite{DBLP:journals/tmis/CiccioM15}, or procedural approaches~\cite{DBLP:conf/sac/LeoniA13}. Translation-based approaches (on automata, or syntax trees) are at the foundation of existing ASP-based solutions~\cite{DBLP:conf/aaai/ChiarielloMP22,DBLP:conf/bpm/KuhlmannCG23}. This paper proposes a novel direct encoding of \declare in ASP that is not based on translations. Moreover, for the first time, we put on common ground (regarding input fact schema) and compare available ASP solutions for conformance checking and query checking. Our experimental evaluation over well-known event logs provides the first aggregate picture of the performance of the methods considered. The results show that our direct encoding, {albeit limited to \declare}, outperforms other ASP-based methods in terms of execution time and peak memory consumption and compares favourably with dedicated libraries. Thus, ASP provides a compact, declarative, and efficient way to implement \declare constraints in the considered tasks. {Interesting future avenues of research are to investigate whether this approach can be extended to \textit{data-aware}~\cite{DBLP:conf/sac/LeoniA13} variants of \declare that take into account data attributes associated to events in a trace, to probabilistic extensions of \declare~\cite{DBLP:conf/iclp/AlvianoIR24,DBLP:conf/pmai/VespaBCLMLC24,DBLP:journals/is/AlmanMMP22} that associate uncertainty (in terms of probabilities) to constraints in a declarative process model, as well as other \declare-based Process Mining tasks}.%

\subsection*{Acknowledgements}
We thank the anonymous reviewers for their careful reading of our manuscript and their many insightful comments and suggestions. 

\medskip

This work was partially supported by the Italian Ministry of Research (MUR) under PRIN project PINPOINT - CUP H23C22000280006; partially by the HypeKG Project of Italian Ministry of Universities and Research (MUR) within Progetti di Rilevante Interesse Nazionale (PRIN) 2022 Program under Grant 2022Y34XNM (CUP: H53D23003710006); partially by the DISTORT Project of Italian Ministry of Universities and Research (MUR) within Progetti di Rilevante Interesse Nazionale Prin 2022 PNRR Program under Grant P2022KHTX7 (CUP: H53D230 08170001) and European Union – Next Generation EU; partially by the PNRR projects FAIR ``Future AI Research'' - Spoke 9 - WP9.1 and WP9.2- CUP H23C22000860006, through the NRRP MUR Program funded by the NextGenerationEU, by the project  "Borgo 4.0", POR Campania FESR 2014-2020, by the French government as part of France 2030 (Grant agreement n°ANR-19-PI3A-0004), the European Union - Next Generation EU as part of the France Relance, and The Italian Ministry of Enterprise Industry and Made in Italy (MIMIT) under project EI-TWIN n. F/310168/05/X56 CUP B29J24000680005.

\paragraph{Competing interests.} The author(s) declare none.

\bibliographystyle{acmtrans}
\bibliography{refs}

\label{lastpage}

\newpage

\appendix 
\section{Validation}
{In this paper, we do not investigate whether there exists an automatic translation from arbitrary \LTLf or \LTLp into ASP rules (beyond indirect techniques), but we focus solely on the \declare constraints, providing ad-hoc encodings written by hand w.r.t their informal semantics. To validate the correctness of these encodings, in our use case, we applied a bounded model checking-like approach, searching for a ``behavioral counterexample" between our encoding of each particular constraint and a \textit{ground truth} logic program - which corresponds to the logic program that captures the behavior of the state machine which is equivalent to the \LTLp definition of the constriant at hand. That is, given a \declare constraint $c$, we consider its \LTLp definition $\varphi_c$ and its corresponding DFA $\mathcal{M}_{\varphi_c}$, and its direct encoding $P_c$. In particular, if we are able to find a trace $\pi$ such that $\pi \models P_c$, but $\pi\not\in \mathcal{L}(\pi)$ - or viceversa, $\pi \not\models P_c$ but $\pi\in \mathcal{L}(\pi)$ - then $\pi$ is a witness of the fact that our direct encoding $P_c$ for the \declare constraint $c$ encodes a wrong behavior (accepting a trace it should not accept, or rejecting a trace it should accept). We tested our encodings for counterexamples of length up to 20 over the $\{a, b, \ast\}$ alphabet, where $a$ plays the role of the constraint activation, $b$ its target, and $\ast$ a placeholder for ``other characters''. As discussed in the automata encoding section, if two propositional symbols do not appear in a \LTLp formula, they are interchangeable in a trace and won't alter the satisfaction of the trace.}

\begin{minipage}{\linewidth}
\centering
\begin{lstlisting}
#const t=20.
time(0..t-1).
activity("*").
activity(A) :- bind(_,_,A).
1 { trace(1,T,A): activity(A) } 1 :- time(T).
:- #count{M: sat(M,C,_)} != 1, constraint(C, _).

\end{lstlisting}
\end{minipage}

{We adapt the conformance checking encoding, adding a (constant) extra term to \texttt{sat/2} predicate (in each encoding involved in the check procedure), to distinguish which evaluation method yields the $sat(\cdot,\cdot)$ atom. That is, instead of $sat(c,tid)$ to model that constraint $c$ holds true over trace $tid$, we use the atoms $sat(automata,c,tid)$, $sat(adhoc,c,tid)$ and $sat(ltlf,c,tid)$ to distinguish satisfiability of the constraint $c$ expressed in the automaton encoding, direct and syntax-tree encoding respectively. 
The above program is to be evaluated together with two distinct conformance checking encoding from Section 3, 4. 
The choice generates the search space for a \LTLp trace. The constraint discards answer sets (e.g., \LTLp traces) that are evaluated \textit{in the same way } by two distinct encodings encodings. That is, it discards traces that are accepted by both encodings or rejected by boths encodings. This logic program yields a model if and only if two \LTLp semantics' encoding are inconsistent with each other. We tested our direct encoding, searching for counterexamples of length up to 20 over the $\{a, b, \ast\}$ against the automaton encoding, and we didn't find any behavioral counterexample.}

\section{Synthetic Log Generation}
{To generate synthetic traces for our experiment, we use the following logic program. Each stable model corresponds to a unique trace. For each constraint in Table~\ref{tab:declare_templates}, we generate a log of 1000 traces (half positive, half negative) over an alphabet of 15 activities. To generate positive traces, we set the external atom $negative$ to false, to generate negative traces we set the $negative$ external atom to true using the Clingo Python API. The constant $t$ represents the length of each trace, it is set as runtime as well. The input facts are the encoding of a Declare constraint, along with the logic program which encodes the semantics (either the automata, syntax tree or direct one). In our experiments, synthetic logs are generated with the direct encoding. To avoid \textit{uninteresting} traces, were constraint are \textit{vacuously} true due to absence of activation/target, we impose that both activation and target should appear at least once.}

\begin{minipage}{\linewidth}
\centering
\begin{lstlisting}
activity("a_0"; ...; "a_14").
#external negative.
#const t=-1.
:- t < 0.
time(0..t-1).

{ trace(0,T,A): activity(A) } = 1 :- time(T).
:- not trace(0,_,A), bind(0, arg_0, A).
:- not trace(0,_,A), bind(0, arg_1, A).
positive :- not negative.
:- not sat(_,0,0), positive.
:- sat(_,0,0), negative.
#show.
#show trace/3.

\end{lstlisting}
\end{minipage}

\section{An example use case for \declare-based process mining}
{%
This is an example application of \declare, regarding the \textit{Sepsis Cases Event Log}~\cite{DBLP:conf/emisa/MannhardtB17}. The log contains events logged by the information system of a dutch hopsital, concerning patients with a diagnosis/suspected diagnosis of sepsis. Here is the official description of the event log:
\textit{This real-life event log contains events of sepsis cases from a hospital. Sepsis is a life threatening condition typically caused by an infection. One case represents the pathway through the hospital. The events were recorded by the ERP (Enterprise Resource Planning) system of the hospital. There are about 1000 cases with in total 15,000 events that were recorded for 16 different activities. Moreover, 39 data attributes are recorded, e.g., the group responsible for the activity, the results of tests and information from checklists. Events and attribute values have been anonymized. The time stamps of events have been randomized, but the time between events within a trace has not been altered.}
Suppose we are interested in conformance checking the log traces against the \declare model containing the constraints:
\begin{itemize}
    \item $Precedence(Antibiotics, IV Liquid)$
    \item $ExclusiveChoice(Release A, Release B)$
    \item $ChainPrecedence(ER Triage, Admission IC)$
\end{itemize}
The model states that the activities $Release A$, $Release B$ (e.g., releasing a patient from the hospital with different types of diagnosis) are mutually exclusive. Admission in the intensive care unit ($Admission IC$) should be immediately preceded by a triage ($ER Triage$); The IV Liquid ($IV Liquid$) exam should not be performed before patients undergo $Antibiotics$.
A \declare model could be either designed by a domain expert, according to e.g. its clinical practice, or clinical guidelines, or discovered automatically from data. Conformance checking allows to identify which traces are compliant, or non-compliant, with the given model, and understand which constraints are violated. To run this example:
$$\texttt{clingo example\_1/*}$$
Suppose instead we are interested in all the activities that immediately follow the admission in intensive care ($Admission IC$) in at most 30\% of the log. This could be useful for clinicians to analyze if clinical practices are correctly being followed. This amounts to performing the query checking task $ChainResponse(AdmissionIC, ?x)$. To run this example:
$$\texttt{clingo example\_2/*}$$
Its output is variable bindings that satisfiy the query checking task. In both commands, the \texttt{filter.lp} file contains only projection rules onto the ``output atoms''.
All the \texttt{*.lp} files are written according to the descriptions in Section 3 and are available at our repository.
More details about the log and other process mining techniques on this log are showcased in ~\cite{DBLP:conf/emisa/MannhardtB17}, although they do not make use of \declare, but non-declarative process mining techniques based on Petri Nets. A detailed application of \declare-based process mining techniques in the healthcare domain is available in ~\cite{DBLP:journals/eswa/RovaniMLA15}, showcasing a case study concerning gastric cancer clinical guidelines and practice.
}

\section{Encodings}
{All the encodings are available under the folders \texttt{ltlf\_base} (syntax tree encoding), \texttt{automata} (automata encoding), and \texttt{asp\_native} (direct encoding) in our repository. Each folder contains a \texttt{templates.lp} file, which are the set of facts that encode the \declare templates as sketched in Section 3 (e.g., automata transition tables and syntax tree reification - notice this file is empty for the direct encoding), and a \texttt{semantics.lp} file which encode logic programs to evaluate automata runs over traces (for the automata encoding), temporal logic operators' semantics as normal rules (for the syntax tree encoding) and the direct translation of \declare into ASP rules for the direct encoding. For completeness, we report here the contents of \texttt{asp\_native\\semantics.lp}, along with pictures that map each constraints' failure conditions to ASP rules.}

For all constraints, a constraint $c$ holds over a trace $tid$ if we derive the $sat(c,tid)$ atom, that is, if we are unable to show a failure condition for $c$ over $tid$:

\begin{lstlisting}
sat(C, TID) :- constraint(C,_), trace(TID,_,_), not fail(C,TID).
\end{lstlisting}

\subsection{Response-based templates}

\subsubsection{Response}\label{appendix:response_picture}

\begin{lstlisting}[caption={$Response$ template},captionpos=b]
witness(C,T,TID) :- 
  constraint(C, "Response"),
  bind(C, arg_0, X), 
  bind(C, arg_1, Y), 
  trace(TID, T, X), trace(TID, T', Y), T' > T.

fail(C,TID) :-
  constraint(C, "Response"),
  bind(C, arg_0, X), 
  bind(C, arg_1, Y), 
  @\textcolor{red}{trace(TID, T, X),}@
  @\textcolor{blue}{not witness(C,T,TID).}@
\end{lstlisting}

\begin{center}
$$\textsf{F}(\textcolor{red}x \land \textcolor{blue}{\lnot \textsf{F} y)}$$
\includesvg[width=\linewidth]{proof_sketch_pictures/failure_response.svg}
\end{center}

\subsubsection{Alternate Response}\label{appendix:alternate_response_picture}

\begin{lstlisting}[caption={$AlternateResponse$ template},captionpos=b]
witness(C,T,TID) :-
  constraint(C, "Alternate Response"),
  bind(C, arg_0, X), 
  bind(C, arg_1, Y), 
  @\textcolor{red}{trace(TID,T,X)}@,
  @\textcolor{blue}{T'' = \#min\{Q: trace(TID,Q,X), Q > T\}}@, 
  @\textcolor{green}{trace(TID,T',Y), T'' > T', T' > T.}@

fail(C,TID) :- 
  constraint(C, "Alternate Response"),
  bind(C, arg_0, X), 
  @\textcolor{red}{trace(TID,T,X)}@, 
  not witness(C,T,TID).
\end{lstlisting}

\begin{center}
$$\textsf{F}(\textcolor{red}{x} \land \textsf{X}_\textsf{w} (\textcolor{blue}{x}\textcolor{green}{ \textsf{R} \lnot y}))$$
\includesvg[width=\linewidth]{proof_sketch_pictures/failure_alternate_response.svg}
\end{center}

Notice that the case where the \textcolor{blue}{x} does not occur is covered by the \texttt{\#min} behavior in \texttt{clingo}, which yields the term \texttt{\#sup} when \texttt{x} does not occur in the trace, hence the arithmetic literal \texttt{T'' > T'} always evaluates to true in this case.

\subsubsection{Chain Response}\label{appendix:chain_response_picture}

\begin{lstlisting}[caption={$ChainResponse$ template},captionpos=b]
fail(C,TID) :- 
  constraint(C, "Chain Response"),
  bind(C, arg_0, X), 
  bind(C, arg_1, Y), 
  @\textcolor{red}{trace(TID,T,X)}@, 
  @\textcolor{blue}{not trace(TID,T+1,Y)}@.
\end{lstlisting}

\begin{center}
$$\textsf{F}(\textcolor{red}{x} \land \textcolor{blue}{\textsf{X}_\textsf{w} \lnot y})$$
\includesvg[width=\linewidth]{proof_sketch_pictures/failure_chain_response.svg}
\end{center}

Here, a single rule covers both failure cases, as $not\ trace(tid,t+1,y)$ literal is true both when the $t+1$ time instant in the trace does not exist, as well as when the $\pi_{i+1} \ne y$.

\subsection{Precedence-based templates}

\subsubsection{Precedence}\label{appendix:precedence_picture}

\begin{lstlisting}[caption={$Precedence$ template},captionpos=b]
fail(C,TID) :-
  constraint(C, "Precedence"),
  bind(C, arg_0, X), 
  bind(C, arg_1, Y), 
  @\textcolor{red}{trace(TID,T',Y)}@,
  @\textcolor{blue}{T = \#min\{Q: trace(TID,Q,X)\}}@,
  @\textcolor{blue}{trace(TID,T,X)}@,
  T' < T.

fail(C,TID) :-
  constraint(C, "Precedence"),
  bind(C, arg_0, X), 
  bind(C, arg_1, Y), 
  @\textcolor{red}{trace(TID,\_,Y)}@,
  @\textcolor{blue}{not trace(TID,\_,X)}.
\end{lstlisting}

We use a dedicated rule to model that \texttt{x} never occurs, when it does not exist, the \texttt{\#min} aggregate would yield the term \texttt{\#sup}.

\begin{center}
$$\textcolor{red}{\textsf{F} y} \land \textcolor{blue}{y \textsf{R} \lnot x}$$
\includesvg[width=\linewidth]{proof_sketch_pictures/failure_precedence.svg}
\end{center}

\subsubsection{Alternate Precedence}\label{appendix:alternate_precedence_picture}

\begin{lstlisting}[caption={$AlternatePrecedence$ template},captionpos=b]
fail(C,TID) :-
  constraint(C, "Alternate Precedence"),
  bind(C, arg_0, X), 
  bind(C, arg_1, Y), 
  @\textcolor{red}{trace(TID,T',Y)}@,
  @\textcolor{blue}{T = \#min\{Q: trace(TID,Q,X)\}}@,
  @\textcolor{blue}{trace(TID,T,X)}@,
  T' < T.

fail(C,TID) :-
  constraint(C, "Alternate Precedence"),
  bind(C, arg_0, X), 
  bind(C, arg_1, Y), 
  @\textcolor{red}{trace(TID,\_,Y)}@,
  @\textcolor{blue}{not trace(TID,\_,X)}@.

fail(C,TID) :-
  constraint(C, "Alternate Precedence"), 
  bind(C, arg_0, X),
  bind(C, arg_1, Y), 
  @\textcolor{purple}{trace(TID, T0, Y)}@, 
  @\textcolor{green}{trace(TID, T2, Y)}@,
  T2 > T0,
  @\textcolor{cyan}{\#count\{Q: trace(TID,Q,X), Q >= T0, Q <= T2\} = 0}.
\end{lstlisting}

\begin{center}
$$(\textcolor{red}{\textsf{F} y} \land \textcolor{blue}{y \textsf{R} \lnot x}) \lor \textsf{F}(\textcolor{purple}{y} \land \textsf{X}(\textcolor{green}{\textsf{F} y} \land \textcolor{green}{y} \textcolor{cyan}{\textsf{R} \lnot x}))$$
\includesvg[width=\linewidth]{proof_sketch_pictures/failure_alternate_precedence.svg}
\end{center}

\subsubsection{Chain Precedence}\label{appendix:chain_precedence_picture}

\begin{lstlisting}[caption={$ChainPrecedence$ template},captionpos=b]
fail(C, TID) :- 
  constraint(C, "Chain Precedence"),
  bind(C, arg_0, X), 
  bind(C, arg_1, Y),
  @\textcolor{red}{trace(TID,T+1,Y)}@,
  trace(TID,T,_),     
  @\textcolor{blue}{not trace(TID,T,X)}@.

fail(C, TID) :-
  constraint(C, "Chain Precedence"), 
  bind(C, arg_1, Y), 
  @\textcolor{green}{trace(TID, 0, Y)}@.
\end{lstlisting}

\begin{center}
$$\textsf{F}(\textcolor{red}{\textsf{X} y} \land \textcolor{blue}{\lnot x}) \lor \textcolor{green}{y}$$
\includesvg[width=\linewidth]{proof_sketch_pictures/failure_chain_precedence.svg}
\end{center}

\subsection{Succession-based templates}

Recall that Succession templates are defined as the conjunction of the corresponding Precedence, Response templates at the same level of the subsumption hierarchy (see Figure~\ref{fig:subsumption_hierarchy}). Hence, its failure conditions are the union of the failure conditions of its subformulae.

\subsubsection{Succession}

\begin{lstlisting}[caption={$Succession$ template. Same failure conditions as $Response$, $Precedence$.},captionpos=b]
fail(C,TID) :-
  constraint(C, "Succession"),
  bind(C, arg_0, X), 
  bind(C, arg_1, Y), 
  trace(TID,T',Y),
  T = #min{Q: trace(TID,Q,X)},
  trace(TID,T,X),
  T' < T.

fail(C,TID) :-
  constraint(C, "Succession"),
  bind(C, arg_0, X), 
  bind(C, arg_1, Y), 
  trace(TID,_,Y),
  not trace(TID,_,X).

witness(C,T,TID) :- 
  constraint(C, "Succession"),
  bind(C, arg_0, X), 
  bind(C, arg_1, Y), 
  trace(TID, T, X), trace(TID, T', Y), T' > T.

fail(C,TID) :-
  constraint(C, "Succession"),
  bind(C, arg_0, X), 
  bind(C, arg_1, Y), 
  trace(TID, T, X), 
  not witness(C,T,TID).
\end{lstlisting}

\subsubsection{Alternate Succession}

\begin{lstlisting}[caption={$AlternateSuccession$ template. Same failure conditions as $AlternateResponse$, $AlternatePrecedence$.},captionpos=b]
witness(C,T,TID) :-
  constraint(C, "Alternate Succession"),
  bind(C, arg_0, X), 
  bind(C, arg_1, Y), 
  trace(TID,T,X),
  T'' = #min{Q: trace(TID,Q,X), Q > T}, 
  trace(TID,T',Y), T'' > T', T' > T.

fail(C,TID) :- 
  constraint(C, "Alternate Succession"),
  bind(C, arg_0, X), 
  trace(TID,T,X), 
  not witness(C,T,TID).

witness(C, T2, TID) :-
  trace(TID, T2, Y),
  T0 = #max{T: trace(TID, T, Y), T2> T},
  trace(TID, T1, X),
  T2 > T1, T1 > T0,
  constraint(C, "Alternate Succession"),
  bind(C, arg_0, X), 
  bind(C, arg_1, Y).

fail(C, TID) :-
  constraint(C, "Alternate Succession"),
  bind(C, arg_0, X),
  bind(C, arg_1, Y),
  trace(TID, T, Y),
  not witness(C, T, TID).

fail(C, TID) :-
  constraint(C, "Alternate Succession"), 
  last(TID, T), 
  bind(C, arg_1, Y), 
  trace(TID, T, Y).
\end{lstlisting}

\subsubsection{Chain Succession}

\begin{lstlisting}[caption={$ChainSuccession$ template. Same failure conditions as $ChainResponse$, $ChainPrecedence$.},captionpos=b]
fail(C,TID) :- 
  constraint(C, "Chain Succession"),
  bind(C, arg_0, X), 
  bind(C, arg_1, Y), 
  trace(TID,T,X), 
  not trace(TID,T+1,Y).

fail(C, TID) :- 
  constraint(C, "Chain Succession"),
  bind(C, arg_0, X), 
  bind(C, arg_1, Y),
  trace(TID,T+1,Y),
  trace(TID,T,_),   
  not trace(TID,T,X).
\end{lstlisting}

\subsection{Choice, Existence templates}

This set of templates, the most general ones in \declare, at the bottom of the subsumption hierarchy, do not involve temporal operators in their \LTLp definition, but only atomic operators (e.g., activity occurrences in the whole trace). They are easily seen as projections on the \texttt{trace/3} predicate.

\begin{lstlisting}[caption={$Choice$ template},captionpos=b]
fail(C, TID) :-
  constraint(C, "Choice"), 
  bind(C, arg_0, X), 
  bind(C, arg_1, Y), 
  trace(TID, _, _), 
  not trace(TID, _, X),
  not trace(TID, _, Y).
\end{lstlisting}

\begin{lstlisting}[caption={$ExclusiveChoice$ template},captionpos=b]
fail(C, TID) :-
  constraint(C, "Exclusive Choice"), 
  bind(C, arg_0, X), 
  bind(C, arg_1, Y), 
  trace(TID,_,X),
  trace(TID,_,Y).

fail(C, TID) :-
  constraint(C, "Exclusive Choice"), 
  bind(C, arg_0, X), 
  bind(C, arg_1, Y), 
  trace(TID, _, _), 
  not trace(TID, _, X), 
  not trace(TID, _, Y).
\end{lstlisting}

\begin{lstlisting}[caption={$RespondedExistence$ template},captionpos=b]
fail(C,TID) :-
  constraint(C, "Responded Existence"),
  bind(C, arg_0, X), 
  bind(C, arg_1, Y), 
  trace(TID,_,X), 
  not trace(TID,_,Y).
\end{lstlisting}

\begin{lstlisting}[caption={$Coexistence$ template},captionpos=b]
fail(C,TID) :-
  constraint(C, "Co-Existence"),
  bind(C, arg_0, X), 
  bind(C, arg_1, Y),
  trace(TID,_,_),
  trace(TID,_,X), 
  not trace(TID,_,Y).

fail(C,TID) :-
  constraint(C, "Co-Existence"),
  bind(C, arg_0, X), 
  bind(C, arg_1, Y),
  trace(TID,_,_),
  trace(TID,_,Y), 
  not trace(TID,_,X).
\end{lstlisting}

\end{document}